\documentclass[aps,pre,twocolumn,groupedaddress,showpacs,amsfonts,10pt,%
floatfix]{revtex4-1}
\usepackage{graphicx}
\usepackage{amsmath}
\usepackage{amssymb}
\usepackage{xcolor}

\let \elastsymbol a

\begin{document}

%Title of paper
\title{Depinning of stiff directed lines in random media}

\author{Horst-Holger Boltz}
\email[]{horst-holger.boltz@udo.edu}

\author{Jan Kierfeld}
\email[]{jan.kierfeld@tu-dortmund.de}

\affiliation{Physics Department, TU Dortmund University, 
44221 Dortmund, Germany}

\date{\today}

\begin{abstract}
 Driven elastic manifolds in random media exhibit a 
 depinning transition  to a state 
 with non-vanishing velocity at a critical driving force. 
We study the depinning of stiff directed lines, which are 
governed by a bending rigidity rather than line tension.
Their equation of motion is 
the (quenched) Herring-Mullins equation, which also 
describes  surface growth governed by surface diffusion. 
Stiff directed lines are particularly interesting as
there is a localization transition in the static problem at a finite
temperature and the commonly exploited time ordering of states by means of
Middleton's theorems (A. Middleton, Phys.\ Rev.\
Lett.\ \textbf{68}, 670 (1992)) is not applicable. We employ
analytical arguments and numerical simulations to determine the critical
exponents and compare our findings with previous works and functional
renormalization group results, which we extend to the 
different line elasticity. We
see evidence for two distinct correlation length exponents.
\end{abstract}

%05.40.-a Fluctuation phenomena, random processes, noise, and Brownian motion
%05.45.-a Nonlinear Dynamics
%68.35.Fx Surface Diffusion

\maketitle

%%%%%%%%%%%%%%%%%%%
\section{Introduction}

Elastic manifolds in random media are one of the most
important model systems in the statistical physics of disordered systems,
which exhibit  disorder-dominated pinned  phases with  many
features common to glassy systems 
\cite{HalpinHealy1995,Blatter1994}.
Likewise, the depinning of an elastic
manifold from a disorder potential under the action of a driving force 
is a paradigm
for  the non-equilibrium dynamical behavior of
disordered systems capturing the avalanche dynamics of many complex
systems if they are driven through a 
complex energy landscape \cite{Paczuski1996}.

In particular, 
the  problem of  a 
 directed line (DL) or directed polymer (i.e., an elastic 
manifold in $1+1$ dimensions) in a random potential and 
driven by a force has been subject of extensive
study \cite{Feigelmann1983,Sneddon1982,Middleton1992,Chauve2000,LeDoussal2002%
,Duemmer2005,LeDoussal2007,Middleton2007,Rosso2007,%
Ferrero2013}. At zero temperature, there is a threshold
force, at which the manifold changes from a localized
state with vanishing mean velocity 
to a moving state with a non-zero mean velocity. 

The depinning transition has been treated within the
 framework of classical critical phenomena 
 by functional renormalization group 
 techniques starting from the more general problem of 
 depinning of $D$-dimensional elastic interfaces (with $D=1$ corresponding 
 to lines). 
 In $D=4-\epsilon$ dimensions,  ``critical'' exponents at depinning 
 can be calculated by functional renormalization 
 using dimensional regularization in an 
  $\epsilon$-expansion  \cite{Narayan1992,Narayan1992b,Nattermann1992}. 

At finite temperature, there is experimental
evidence for a creep motion at any non-vanishing driving forces which
can be understood qualitatively as thermally activated crossing of energy
barriers which result from an interplay of
 both elastic energies of the line and the disorder potential.

\begin{figure}
 \includegraphics[width=0.95\linewidth]{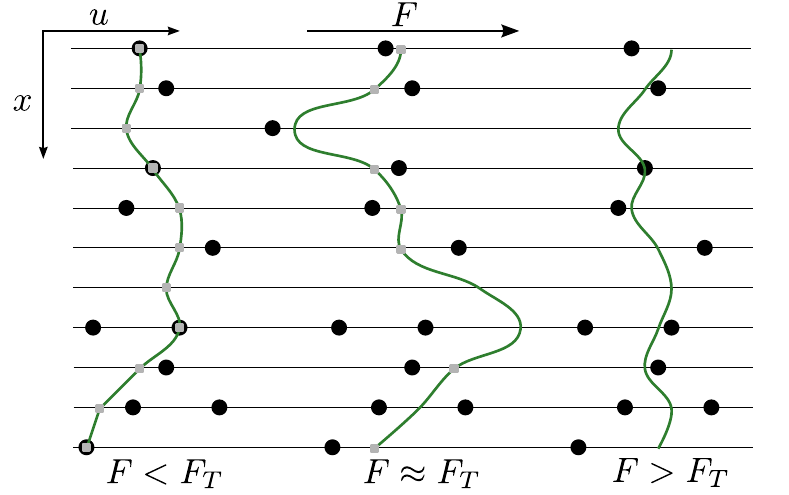}
 \caption{
  Schematic figure of typical SDL configurations 
  below, at, and above the depinning transition (from left to right)
in random-potential (RB)
  disorder. The black circles represent especially favorable locations
within the random potential. Segments of the line that are currently not moving
are marked with gray squares.}
\label{fig:bild}
\end{figure}

The energy of DLs such as flux lines, domain walls, wetting fronts 
is proportional to their length; therefore, the elastic properties
of directed lines are governed by their line tension,
which  favors the straight 
configuration of shortest length. 
Here, we concentrate on  stiff
directed lines (SDLs), whose elastic energy is given by the curvature of the
line and, thus, represents a bending energy. 
  This gives rise to configurations 
 which are locally curvature-free and straight but, in contrast to the 
 DL,   the straight segments of SDLs 
 can assume any orientation even if this increases
 the total length of the line.

There are a number of applications and interesting 
general theoretical issues concerning SDLs in random media. 
The  overdamped equation of motion of a SDL 
is the (fourth-order) Herring-Mullins linear diffusion
equation \cite{Herring1951,Mullins1957}, 
which also describes surface growth 
governed by surface diffusion. 
The depinning dynamics of the Herring-Mullins equation in quenched disorder
has been subject of a number of 
prior studies \cite{Park2000,Lee2006,Liu2008,Song2006}, 
whose findings (e.g., an unphysically small roughness exponent)
differ in part significantly from ours as we will point out below
(see Sec.\ \ref{sec:prev}). 
SDLs also describe  semiflexible polymers with contour lengths 
smaller than their persistence length for bending fluctuations, 
such that the assumption of a directed line 
is not violated \cite{Kierfeld2006}. 
Our results can be applied to 
the depinning dynamics of 
 semiflexible polymers such as DNA or cytoskeletal filaments like F-actin
 in a random environment, such as   a porous
 medium, as long as the correlation length of the 
depinning transition is smaller than the persistence length. 
As in other semiflexible polymer phase transitions (such as adsorption)
non-universal quantities such as the value of the 
depinning threshold itself will be governed by the bending elasticity. 
At the depinning transition, where
the correlation length diverges, semiflexible polymers will exhibit a
crossover to critical properties of effectively flexible lines (DLs)
with a segment length set by the persistence length.

Moreover,  static 
SDLs in a random potential feature a disorder-driven 
localization transition  at finite temperatures
already in $1+1$ dimensions \cite{Bundschuh2002,Boltz2012,Boltz2013}.
 Due to an interesting 
dimensional shift in the problem, an analogous transition occurs 
for DLs only in higher dimensions. 
In principle, this offers
 an opportunity to observe new phenomena arising from an 
interplay of depinning and delocalization for SDLs (the disorder
used in Refs.\ \cite{Park2000,Lee2006,Liu2008,Song2006} 
does not feature such a
transition in the static problem).
The localization transition at a finite temperature also offers the
opportunity to test the usage of static quantities 
in the treatment of creep
motion, because the SDL is not pinned by disorder above the critical
temperature.

A lot of the progress for the depinning theory of DLs 
has been based on two basic 
theorems due to Middleton \cite{Middleton1992},
which essentially state that a forward moving DL 
can only move forward and will
stop in a localized configuration, if such exists.
This allows for an
 unambiguous time-ordering of a sequence of states. 
We will show that these
theorems do not hold for the SDL, which can be seen as a consequence of the
next-to-neighbor terms that are introduced by the bending elasticity.

The paper is organized as follows. We present the model and the
relevant equations of motion in Sec.\ \ref{sec:model}, where we also comment
on some equilibrium properties and previous work on the SDL. 
In Sec.\
\ref{sec:ana} we present analytical results based on 
 scaling arguments and  functional renormalization
 group calculations. 
In
Sec.\ \ref{sec:num} we present numerical methods and results and, in Sec.\
\ref{sec:middle}, we comment on the non-applicability 
of Middleton's theorems for SDLs.
We conclude with a summary of our results in Sec.\
\ref{sec:concl}.

%%%%%%%%%%%%%%%%%%%%%%%%%%
\section{Model} \label{sec:model}

A general approach to  driven 
elastic manifolds starts from  
$1+D$ dimensional manifolds with an  elastic energy
\begin{align}
 \mathcal H_{\text{el}} &= \frac{1}{2}\int \mathrm{d}^{D}x \left(
\partial^\elastsymbol_x u\right)^2 \text{.} \label{eq:ham}
\end{align}
As an external pulling force $F$  intrinsically couples 
to only one transversal displacement component $u$, there is no
substantial gain in treating more than one transverse dimension,
 and we will
restrict our analysis to this case. We assume that
the manifold's internal coordinates $x$ are bounded to the $D$-dimensional
hypercube $[0,L]^D$ and call $L$ the system size (or length in $D=1$). We are
mostly interested in the simplest case of lines with $D=1$. 
The order of the derivative $\elastsymbol$ distinguishes
different
kind of elasticities, $\elastsymbol=1$ is the directed line (DL)
with tensional elasticity,
and $\elastsymbol=2$ is the stiff directed
line \cite{Boltz2012,Boltz2013} (SDL) with bending elasticity.

The equilibrium statistics of the SDL and DL model are related as there is a
mapping of the SDL to the DL model in higher transverse
dimensions in problems with short-ranged
potentials \cite{Bundschuh2000,Kierfeld2005}. In an earlier work we extended
this mapping between SDL and DL in short-ranged random
potentials \cite{Boltz2012,Boltz2013}.
In the context of polymers, the SDL model is often used as a weak-bending 
approximation to the so-called worm-like chain or Kratky-Porod
model \cite{Harris1966,Kratky1949}, which is the basic model for inextensible
semiflexible polymers, such as DNA, cytoskeletal filaments like F-actin,
or polyelectrolytes. The 
weak-bending approximation is only applicable on length 
scales below the persistence length $L_p$, where tangent fluctuations 
$\overline{\langle (\partial_x u)^2 \rangle} <1$ remain small.
The persistence length contains thermal and disorder contributions
as discussed in Refs.\ \cite{Boltz2012,Boltz2013}.
Additionally, there is a relation of the SDL model to
surface growth models, on which we will comment below.

In suitable units, the overdamped equation  of an elastic 
line with an elastic energy $\mathcal H_{\text{el}}$ as in \eqref{eq:ham}
can be written as 
\begin{align}
 \frac{\partial u(x)}{\partial t} &= - \frac{\delta \mathcal
   H_{\text{el}}}{\delta u(x)} + \eta(x,u)+ F
\label{eq:EOM}
\end{align}
at zero temperature. 
The first term on the right hand side represents the elastic 
forces as obtained from variation of the elastic energy \eqref{eq:ham}.
 The force  $F$ denotes a static, 
uniform pulling force, which
tends to depin the line from the disordered medium,  and
$\eta(x,u)$ is a quenched force  due to the disordered medium. 
For this quenched force we distinguish two cases in the following,
random-field  and random-potential disorder. 
For random-field
(RF) disorder, $\eta(x,u)$ is a random variable 
 with zero mean and short-ranged correlations
\begin{align}
 \overline{\eta(x,u)\eta(x',u')} &\propto \delta(x-x') \delta(u-u') \quad
\text{(RF),}
\end{align}
whereas for random-potential or random-bond (RB) disorder the
force $\eta(x,u)=-\partial_u V(x,u)$ stems from a random potential $V(x,u)$,
which features zero
mean and short-ranged correlations. 
Hence, after a Fourier transformation in the
transverse dimension, we can write
\begin{align}
 \overline{\eta(x,q)\eta(x',q')} &\propto q^2 \delta(x-x') \delta(q+q') \quad
\text{(RB).}
\end{align}
Within this work we will mostly focus on random-potential (RB) 
disorder; in particular, all our numerical results  in
Sec.\ \ref{sec:num} are for RB disorder. 
The analytical arguments in Sec.\ \ref{sec:ana}, i.e., scaling 
relations and functional renormalization group results will be 
applied to both types of disorder. 

If  thermal fluctuations at temperature $T$ are included, 
 an additional time-dependent white noise
$\tau(x,t)$ with zero mean and correlations
\begin{align}
 \langle \tau(x,t) \tau(x',t')\rangle_T &= 2 T \delta(x-x') \delta(t-t')
\end{align}
is added on the right hand side of eq.\ \eqref{eq:EOM}.
We use $\langle X \rangle$ to denote averages over time,
$[ X ]$ for spatial averages at a given time, $\overline{X}$ for the average
over realizations of the quenched disorder and $\langle X\rangle_T$ for
the average over realizations of the thermal noise. 
A subscript $c$ at an average denotes a cumulant.

At zero temperature, there is a finite threshold
value $F_T$ at which the velocity
\begin{align}
 v(t)&=\overline{[\dot{u}(x,t)]}
\end{align}
of a driven elastic manifold in a pinning potential 
becomes nonzero  in the
limit of very large times $t$
\begin{align}
 \langle v(t)\rangle&\begin{cases}=0 & F\leq F_T
 \\>0 &  F > F_T\end{cases} \text{.}
\end{align}

The  shape of the driven line and 
its dynamics are usually characterized
 by  the roughness exponent $\zeta$ and the 
dynamical exponent $z$, which describe how the
roughness or width of the line, 
\begin{align}
 w^2(t) &= \overline{[ u(x,t)^2-[u(x,t)]^2]} = \overline{[ u(x,t)^2 ]_c},
\label{eq:defw}
\end{align}
scales with the system size $L$
and the time $t$:
\begin{align}
 w^2(t) \sim \begin{cases}  t^{2\zeta/z} & t\ll t_L \\
   L^{2\zeta} & t\gg t_L\end{cases} 
\label{eq:defzetaz}
\end{align}
with the  typical time scale 
\begin{equation}
  t_L \sim L^z.
\label{eq:defz}
\end{equation}

SDLs  with $\elastsymbol=2$ 
 are closely related to surface growth models for molecular
beam epitaxy (MBE) \cite{Barabasi1995}. 
In the presence of surface diffusion,
MBE has been described by a (quenched) 
Herring-Mullins linear diffusion
equation \cite{Herring1951,Mullins1957,DasSarma1994, Racz1994}
for a surface described by a height profile $u(x,t)$, 
\begin{align}
 \frac{\partial u(x)}{\partial t} &= - \nabla^4 u(x) + \eta(x,u)+ F
\text{.}
\label{eq:mulher}
\end{align} 
In the Herring-Mullins limit, it is assumed that effects from a 
surface tension can be neglected as compared to surface diffusion effects. 
Surface tension would give rise to additional $\nabla^2u$-terms.
In this context, the quantity  $F$ describes the constant flux of particles 
onto the surface and $\eta(x,u)$ random fluctuations in the deposition 
process.

In suitable units\footnote{With a full set of parameters the equation would
read $ \lambda {\partial u(x)}/{\partial t} = - \varkappa \nabla^4 u(x) + g
\eta(x,u)+ F$, which reduces to the given form by rescaling $x = x/x_0$,
$u=u/u_0$ and $t=t/t_0$ with $x_0$ given by the discretization, $u_0 =
(g/\varkappa)^{2/3} x_0^3$ and $t_0 = \lambda x_0^4/\varkappa$.}, the
Herring-Mullins equation (\ref{eq:mulher}) is 
 equivalent to the overdamped equation
of motion \eqref{eq:EOM} of the  SDL. 
Without external
forces ($\eta=0$, $F=0$) the exponent $\zeta$ and $z$ 
take their thermal values $\zeta_{th}=3/2$ and $z_{th}=4$ \cite{Barabasi1995}. 
 The observation of the
``super-rough'' $\zeta=3/2>1$ in tumor cells \cite{Bru1998} hints towards
further experimental relevance of the Herring-Mullins equation.

For the DL  with $\elastsymbol=1$,  the equation of motion is
the quenched Edwards-Wilkinson  equation \cite{Edwards1982}
\begin{align}
 \frac{\partial u(x)}{\partial t} &= \nabla^2 u(x) + \eta(x,u)+ F \text{,}
\end{align} 
and the thermal exponents are $\zeta_{th}=1/2$ and $z_{th}=2$. 
Generally, the thermal
exponents are given by $\zeta_{th}=\elastsymbol-\frac{D}{2}$ and
$z_{th}=2\elastsymbol$.

%%%%%%%%%%%%%%%%%%%%%%%%%
\subsection{Equilibrium properties}

The equilibrium ($F=0$) problem of a SDL in a $1+1$-dimensional medium 
with RB disorder features a localization 
transition at a
finite temperature $T_c$ as we pointed out in Refs.\ 
\cite{Boltz2012,Boltz2013}. 
The roughness exponent is
$\zeta_\text{eq,RB}\approx1.59>3/2$ in the disorder-dominated phase 
for $T<T_c$ and assumes the thermal value $\zeta=3/2$
for $T>T_c$.
In contrast to the SDL, 
the DL with one transverse dimension is localized 
for all temperatures with a
roughness exponent $\zeta_\text{eq,RB}=2/3$ \cite{Kardar1987}. 
This implies that the
SDL in RB disorder 
 offers the opportunity to study the dynamics of 
an unlocalized elastic
manifold in disorder for $T>T_c$ and the interplay of 
the delocalization transition at $T=T_c$ and a depinning transition 
at $F=F_T$.

For RF disorder, functional renormalization 
group approaches \cite{Chauve2001,LeDoussal2004}
using  an  expansion in  $\varepsilon=4\elastsymbol-D$ around 
 the upper critical dimension
$D_c=4\elastsymbol$
give a static
 roughness exponent $\zeta_\text{eq,RF}=\varepsilon/3$
to at least two (and possibly all) orders in 
an expansion in  $\varepsilon$ and in good agreement with
numerical results both for the DL \cite{Belanger1991,Seppala1998}
and the SDL \cite{Lee2000}. 
In a
 discrete model that directly implemented surface diffusion and was proposed
 to correspond to the undriven quenched Herring-Mullins equation a
 differing exponent $\zeta_\text{eq,RF}\approx 1.93$ 
 was found for the SDL \cite{Park2000}.

The result  $\zeta_\text{eq,RF}=\varepsilon/3$ is the
simple scaling or ``Flory'' result, that follows from balancing the typical
elastic energy of a line with displacement $u$, which scales as
$E_{\text{el}}\sim L^D (u/L^\elastsymbol)^2$, with the typical disorder energy
$E_{\text{dis}}\sim \sqrt{L^D u}$ as the disorder energy is picked up at $L^D$
independent sites and its correlator decreases linearly in $u$
for large enough $u$ \cite{Fisher1986}.
Similar arguments fail to reproduce the non-trivial RB
roughness exponent, but can provide bounds to it as 
discussed in Refs.\ \cite{Boltz2012,Boltz2013}.

%%%%%%%%%%%%%%%%%%%
\subsection{Previous work on the depinning of the 
quenched Herring-Mullins equation}\label{sec:prev}

There has been some previous 
work on the depinning of SDLs with RF disorder. From
renormalization group analysis it is expected 
that the critical exponents of the
depinning transition are universal for all disorders with shorter ranged
correlations than RF (including RB), although a different scenario is 
possible in principle \cite{LeDoussal2002}. 
For the DL the exponents do
coincide for RF and RB disorder \cite{Ferrero2013}.

The roughness exponents previously found at  the depinning of a SDL in 
RF disorder are $\zeta \approx 1.48-1.50$ and
$\zeta =1.48$ and a dynamical exponent $z\approx 1.77-1.78$ 
\cite{Lee2006,Liu2008}. Furthermore, in a discrete model \cite{Song2006} 
based on
the quenched Herring-Mullins equation  $\zeta =1.35$ and $z=1.60$ 
have been found at depinning.
One obvious problem with these values for the 
roughness exponent $\zeta$ is that they
are {\em smaller} than the 
thermal value $\zeta_{th}=3/2$, i.e., that disorder decreases 
the roughness of the line.  We will comment below in more detail 
on similarities and differences in the findings of these studies
to ours.

%%%%%%%%%%%%%%%%%%%%%%%%%%%%%5
\section{Analytical results} \label{sec:ana}

%%%%%%%%%%%%%%%%%%%%%%%%%%%%%%%%%%
\subsection{Critical exponents and scaling relations} \label{sec:scaling}

In order to describe the  depinning of driven elastic lines in a random
medium within the framework of classic critical phenomena
\cite{Narayan1992,Narayan1992b,Nattermann1992,Narayan1993},
the roughness exponent $\zeta$ and dynamical exponent $z$ 
introduced in eq.\ (\ref{eq:defzetaz}) are not sufficient 
but one additional  exponent related to the control parameter, 
the driving  force $F$, is needed.
In the vicinity of the depinning threshold $F_T$
we can introduce two exponents describing the  
``order parameter'', which is 
the velocity $v$ of the center of mass, and the correlation length $\xi$:
\begin{align}
 v&\sim (F-F_T)^{\beta}
 \label{eq:defbeta}\\
 \xi&\sim (F-F_T)^{-\nu}.
\label{eq:defnu}
\end{align}
The correlation length $\xi$ gives the typical length 
of segments that rearrange during the avalanche-like
motion close to the threshold;
 the typical time scale for this 
segment motion is $t_\xi \sim \xi^z$.

We can use one of these exponents, e.g.\ 
the correlation length exponent $\nu$, to 
obtain from 
the equilibrium scaling relation 
\begin{equation}
   w(t,L)  =   t^{\zeta/z} g(t/t_L^z)
\end{equation}
with a scaling function $g(x)$ (with 
$g(0) \approx 1$ and $g(x)\sim x^{-\zeta/z}$ for $x\gg 1$),
which underlies eq.\ (\ref{eq:defzetaz}), 
a corresponding scaling relation close to depinning, 
\begin{equation}
   w(t,F)  =   t^{\zeta/z} g(t/t_\xi^z) = 
            t^{\zeta/z} f_{\pm}(t^{1/\nu z} (F-F_T)),
\label{eq:scalingwtF}
\end{equation}
with   scaling functions $f_{\pm}(x)$ (for forces 
above (+) and below (-) the threshold and with 
$f_{\pm}(0) \approx 1$ and $f_{\pm}(x)\sim |x|^{-\zeta \nu}$ for $|x|\gg 1$).

There are two scaling laws relating the 
exponents $\beta$ and $\nu$  to  
the roughness exponent $\zeta$
and the dynamical exponent $z$ at the depinning transition. 
The first scaling law simply establishes a relation between 
$\beta$ and $\nu$ using that  $v\sim w(t_\xi)/t_\xi\sim \xi^{\zeta-z}$
(for  $t\gg t_\xi$), which results in 
\begin{equation}
 \nu = \frac{\beta}{z-\zeta}.
 \label{eq:betascalrel}
\end{equation}
This relation 
is valid independently of the form of the elastic energy, i.e., 
independent of $\elastsymbol$. As all
exponents should be positive this also implies
 $z>\zeta$.
The other relation comes from an 
additional tilt symmetry of the equation of motion
\cite{Nattermann1992,Narayan1993}, which   leads  to
\begin{align}
 \nu &= \frac{1}{4-\zeta}
\label{eq:nuscalrel}
\end{align}
for the SDL ($\elastsymbol=2$) 
or, for general elasticity, to $\nu=(2\elastsymbol-\zeta)^{-1}$.
The  relations (\ref{eq:betascalrel}) and (\ref{eq:nuscalrel}) 
should hold both for RF and RB disorder at depinning.

For the analysis of simulation data it is convenient to infer 
exponents from the short time scaling properties of the 
velocity, which follows from the scaling (\ref{eq:scalingwtF})
and $v\sim w/t$:
\begin{align}
 v(t,F)  &\sim t^{\zeta/z-1} f_{\pm}(t^{1/\nu z} (F-F_T))
 \nonumber\\
     &\sim t^{-\delta}f_{\pm}(t^{\gamma} (F-F_T))    
\label{eq:vf}
\end{align}
where we introduce two auxiliary exponents 
\begin{align}
\gamma &=1/\nu z
\label{eq:gamma}\\
 \delta &=1-\zeta/z  = \beta/\nu z= \beta\gamma,
   \label{eq:delta}
\end{align}
for  convenient data analysis [using the  scaling relation
(\ref{eq:betascalrel})  in eq.\ (\ref{eq:delta})].

The exponent values  obtained previously  in Ref.\ \cite{Lee2006}
are $\zeta \approx 1.50$ and $\nu \approx 1.01$ at the SDL
depinning transition (for RF disorder).
These values are problematic as they violate 
the scaling relation (\ref{eq:nuscalrel}).
One reason for this problem might be that 
the exponent  $\zeta$ 
has been determined by direct measurement of the roughness $w(L)$
and its scaling for different system sizes $L$. 
However, such an approach  is strongly influenced by the 
choice of the transverse system size
(which should be $M\sim L^\zeta$) because the value for the 
critical force $F_T$ depends also on the transverse system size. 
As Ref.\ \cite{Lee2006}
contains two other independently  measured exponents, 
namely $\delta$ and $\beta$ in our nomenclature, and
scaling relations  (\ref{eq:betascalrel}) and (\ref{eq:nuscalrel}) 
imply 
$\zeta = 4\beta(1-\delta)/(\delta + \beta(1-\delta))$. 
 we can give a resulting ``scaling'' roughness
exponent $\zeta_{\text{scaling}}\approx 2.4$, which strongly differs.

There is a another exponent that is often referred to as $\nu$ or
$\nu_{\text{FS}}$ describing
the scaling of the sample-to-sample fluctuations of the threshold force
\begin{equation}
 \Delta F_T \sim L^{-1/\nu_\text{FS}} 
\label{eq:nuFS}
\end{equation}
in a system of finite size $L$.
In general, $\nu$ and $\nu_\text{FS}$ do not have to coincide. 
For the DL, $\nu=\nu_\text{FS}\approx 4/3$ has been
confirmed \cite{Duemmer2005,Bolech2004}, whereas for the charge 
density wave problem
(periodic potential), $\nu$ and $\nu_\text{FS}$ are
distinct \cite{Middleton1992b}.

This might affect  the scaling relations (\ref{eq:betascalrel})
and the auxiliary exponents $\gamma$ and $\delta$, 
see eqs.\ (\ref{eq:gamma}) and (\ref{eq:delta}), which could read 
$\nu_\text{FS} = \beta/(z-\zeta)$, $\gamma= 1/\nu_\text{FS} z$, and 
$\delta = \beta/\nu_\text{FS} z$.
This happens if threshold force fluctuations by 
sample-to-sample disorder fluctuations on a scale $L$, 
$\Delta F_T\sim L^{-1/\nu_\text{FS}}$, 
are larger than the excess  to the threshold force necessary 
to depin a segment of length $L$, $F-F_T \sim L^{-1/\nu}$, see eq.\ 
(\ref{eq:defnu}). Therefore, we expect  $\nu<\nu_\text{FS}$ 
if  $\nu$ and $\nu_\text{FS}$ are distinct.

Fluctuations in the depinning force $F_T$ origin from the fluctuations in the
disorder. A finite manifold of size $L$ and width $w\sim L^\zeta$ 
occupying a volume $L^Dw \sim L^{\zeta+D}$ should at least pick up the same 
free energy fluctuations as a 
summation of i.i.d.\ random numbers which are 
$\sim (\text{Vol.})^{\frac{1}{2}}$. This results in  a general 
lower limit for  the 
depinning force fluctuations \cite{Chayes1986}
\begin{equation}
 \Delta F_T \geq c (\text{Vol.})^{-\frac{1}{2}} 
= \tilde{c} L^{-\frac{\zeta+D}{2}}
\end{equation}
thus giving
\begin{equation}
 \nu_\text{FS} \geq \frac{2}{\zeta+D}.
  \label{eq:nufsbound}
\end{equation}
It has been argued that $\nu=\nu_\text{FS}$ for an elastic
line as long as the line continuously ``explores'' new regions of the
disorder \cite{Narayan1993}. In this interpretation, distinct
correlation length exponents $\nu$ and 
$\nu_\text{FS}$ for the charge density wave are a 
manifestation of the fact that the line ``knows'' the total potential
 at each point due to its periodicity. 
If $\nu=\nu_\text{FS}$ holds, the bound
(\ref{eq:nufsbound}) is equivalent to a lower bound to the 
roughness exponent $\zeta$ at depinning,
\begin{align}
 \zeta \geq \varepsilon/3 \quad \text{ if } \nu=\nu_\text{FS},
\label{eq:zetafsbound}
\end{align}
which is valid for all elastic energies of the form (\ref{eq:ham}),
i.e., for all $\elastsymbol$. 
The contraposition is equally important: if the
roughness is less than
$\varepsilon/3$, this implies that $\nu$ and $\nu_\text{FS}$ are distinct.

An
upper bound to the roughness exponent comes from studying the line in the
Larkin approximation with a constant ($u$-independent) random force 
 acting  on every segment of the
line \cite{Larkin1970,Blatter1994}. 
The resulting Larkin roughness exponent is 
$\zeta_{\text{Larkin}}=\varepsilon/2$. As the line can gather unbound energy via
large undulations in accordance with the force  this represents  an
upper bound to the problem with finite potential range and, therefore,
\begin{align}
 \zeta \leq \varepsilon/2 \text{,}
\end{align}
which holds for the roughness exponents $\zeta$ 
below, at,  and above depinning.

%%%%%%%%%%%%%%%%%%%%%%%%%%%%%%%%%%%%%%%%%%%%%%%
\subsection{Functional renormalization group}
\label{sec:frg}

\begin{figure}
 \includegraphics[width=0.95\linewidth]{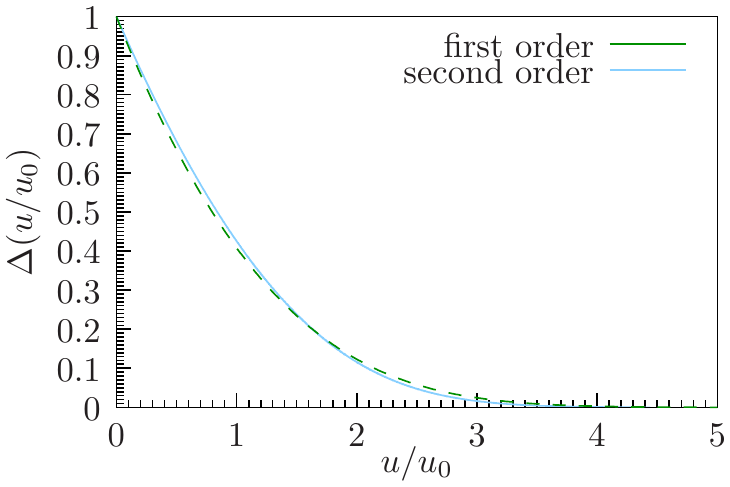}
 \caption{
The fixed point solution to the 
functional renormalization group flow equation of
Ref.\ \cite{LeDoussal2002} for the disorder correlator $\Delta(u)$.
The length scale $u_0$ is determined 
via $\int\mathrm{d}u \Delta{(u/u_0)} = 1$.
The two-loop solution corresponds to a roughness
exponent given by eq.\ \eqref{eq:frg}. The small deviations between  first and
second order  give rise to distinct 
roughness exponents (\ref{eq:zeta12}).
}
 \label{fig:frg}
\end{figure}

It has originally  been suggested that the roughness exponent
$\zeta$ at the threshold force $F=F_T$ 
is independent of the type of disorder (RB or RF) and 
 coincides  with the static roughness of
the line in a medium with RF disorder 
$\zeta_\text{eq,RF}=\varepsilon/3$
to all orders of $\varepsilon$ \cite{Narayan1993}. 
The discrepancy to numerical
simulations \cite{Leschhorn1993} has been solved by means of the two-loop
functional renormalization group (FRG) \cite{LeDoussal2002},
which gives for the roughness exponent at depinning both 
for RB and RF disorder
\begin{align}
 \zeta^{(2\elastsymbol)}&=\frac{\varepsilon}{3} 
 + \frac{X^{(2\elastsymbol)}\varepsilon^2}{27\sqrt{2}\gamma} 
   + \mathcal O (\varepsilon^3) 
\label{eq:frg}
\end{align}
with the Euler-Mascheroni constant $\gamma$ and a constant $X^{(2\elastsymbol)}$
that
depends on the form of the elastic energy, especially $X^{(2)} = 1$ and
$X^{(4)}=-1/6$. The FRG approach is based on a flow equation for  
 the disorder force correlator
$\Delta(u)$ defined by 
\begin{align}
  \overline{\eta(x,u)\eta(x',u')} &= \delta(x-x') \Delta(u-u') \text{.}
\end{align}
At two-loop order the FRG flow converges to the same 
fixed point disorder correlator (shown in Fig.\ \ref{fig:frg}) 
both for RB and RF disorder. Therefore, two-loop FRG predicts 
identical roughness exponents at depinning. 

As $\varepsilon=4\elastsymbol-d$ is rather large for the SDL,
 the two-loop contribution is important. This can be seen in the
significant deviation of the two naive results 
\begin{equation}
 \zeta_{\text{1-loop}}\approx 2.33~~\mbox{and}~~
 \zeta_{\text{2-loop}}\approx 1.94 
\label{eq:zeta12}
\end{equation}
using a direct evaluation of
\eqref{eq:frg} (Pad\'{e} approximants give values
$\zeta_{\text{2-loop}} \approx 1.9432$ to
$\zeta_{\text{2-loop}}\approx 1.999$).
 Thus the SDL roughness at depinning is expected to be above the
static value for zero external force \cite{Boltz2012,Boltz2013}
$\zeta_{eq,RB}\approx1.59$ for RB
disorder, but below the static result 
$\zeta_{eq,RF}=7/3$ for RF disorder.
We note that any extrapolation of the results in eq.\
(\ref{eq:frg}) necessarily violates the bound (\ref{eq:zetafsbound}),
which holds if $\nu=\nu_\text{FS}$, as the
negative two-loop contribution leads to $\zeta^{(4)}<\frac{\varepsilon}{3}$ for
some finite $\varepsilon$.
This is an indication for two distinct exponents 
$\nu$ and $\nu_\text{FS}$.

In Fig.\ \ref{fig:frg} we show the numeric solution of the 
FRG fixed point for the disorder correlator $\Delta(u)$ 
 verifying that eq.\ (\ref{eq:frg}) does indeed 
correspond to the unique
non-negative faster than algebraically decaying (convex in double logarithmic
plot) solution. We followed the
numerical procedure outlined in Refs.\ \cite{LeDoussal2002, LeDoussal2004}. 
The new second order contribution $y_2$ giving the two-loop
contribution in the Ansatz $\Delta(u)=\varepsilon/3  y_1(x)+\varepsilon^2/18
y_2(x)+O(\varepsilon^3)$ can be approximated by the Taylor series
\begin{align}
y_2(x)&\approx 0.190021 u-0.1613 u^2+3.37491 \times 10^{-2} u^3
\nonumber\\&+3.21649\times 10^{-3} u^4-4.32055 \times 10^{-4}
u^5\nonumber\\&-2.55032\times	 10^{-4} u^6 -4.7737 \times 10^{-5}
u^7\nonumber 
\\&+1.0426\times 
10^{-6} u^8+3.44412\times 10^{-6} u^9+O(u^{10}) \text{.}
\label{eq:frg2loop}
\end{align}

The FRG calculation presented in Ref.\
\cite{LeDoussal2002} is in principle also capable of determining the dynamical
exponent $z$ and, thus, all exponents. 
However, to two loops this involves the
evaluation (to leading order in $\varepsilon$) of the 
``correction to friction''
which remains an open task. It is possible (details are given in appendix
\ref{app:bounds}) to find bounds 
for the value of the dynamical exponent $z$ in two-loop order
\begin{align}
 1.86 \leq z_{\text{2-loop}} \leq 2.31 \quad \text{.}
\label{eq:zbounds}
\end{align}
  The exponents at
one-loop order for the SDL are
given in Table \ref{tab:exp} together with the previous numerical findings. 

\begin{table}
\begin{tabular}{l|ccccc}
 \toprule
 exponent & $\zeta$ & $z$ & $\nu$ & $\beta$ & $\delta$ \\ \hline
 FRG & $7/3$   & $22/9$ & $3/5$ & $1/15$ & $1/22$ \\
 simulation Ref.\ \cite{Lee2006} 
  & $1.50$ & $1.78$ & $1.01$ & $0.289(8)$ & $0.160(5)$\\
 simulation this paper & $2.00$ & $2.15$ & $0.50$ & $0.29$ &
$0.07$\\\botrule
\end{tabular}
\caption{
Critical exponents at the SDL depinning transition determined via one-loop
functional renormalization group (FRG) \cite{LeDoussal2002} 
and numerical simulation results from Ref.\ \cite{Lee2006} and 
this work.
The last line uses the numerical values for
 $\beta$ and $\delta$ and
the scaling relations of Sec.\ \ref{sec:scaling}.
}
\label{tab:exp}
\end{table}

An interesting question in the FRG  analysis is the stability
of the fixed point solution. It has been argued
\cite{LeDoussal2002,Narayan1993} that the two previously defined correlation
length exponents coincide, that is $\nu=\nu_\text{FS}$, if the fixed point
solution for the disorder correlator is stable. This is in agreement with the
results for charge density waves (fixed point unstable
\cite{Narayan1992,Narayan1992b}, $\nu\neq\nu_\text{FS}$ \cite{Middleton1992b})
and the DL (fixed point presumably stable \cite{LeDoussal2002},
$\nu\approx\nu_\text{FS}$ \cite{Duemmer2005}). 
We did not try to perform a full
stability analysis, but we note that the simple argument of 
Ref.\ \cite{LeDoussal2002} for
the instability of the fixed point for charge density waves might 
also hold for SDLs: after integrating the 
FRG flow equation from $u=0^+$ to $u=\infty$ it reads
\begin{align}
 -m \partial_m \int_0^\infty\! \Delta(u) \mathrm{d}u &= (\varepsilon -
3\zeta)\int_0^\infty\! \Delta(u)\mathrm{d}u - X^{(4)} \Delta'(0^+)^3
\text{.} 
\label{eq:fp}
\end{align}
The second contribution on the right hand side 
is negative because (to two loops) 
\begin{align}
 \zeta^{(2\elastsymbol)} &= \frac{1}{3}\varepsilon - \frac{X^{(2\elastsymbol)}
\Delta'(0^+)^3}{3 \int \Delta} = \frac{\varepsilon}{3} +
\zeta^{(2\elastsymbol)}_2 \varepsilon^2 
\label{eq:zeta}
\end{align}
and $\zeta^{(4)}_2<0$ (with the shorter notation $\int \Delta\equiv\int_0^\infty
\mathrm{d}u \Delta(u)$). More importantly, this implies that
$\zeta<\varepsilon/3$ and, thus, the FRG fixed point  of $\int \Delta$ 
is unstable. 
The instability of the fixed 
point of eq.\ \eqref{eq:fp} leads to a flow of the form
\begin{align}
 \Delta_m(u)=\Delta^*(u)+cm^{-(\varepsilon-3\zeta)}
\end{align}
with $c= m_0^{\varepsilon-3\zeta} \int_0^\infty (\Delta_{m_0}(u)-\Delta^*(u) )
\mathrm{d}u$. Thus an additional constant ($u$-independent) 
contribution to the fixed point $\Delta^*(u)$ is generated,
 that grows as $m$ goes to zero if $\zeta<\varepsilon/3$. This means that to
two-loop order a random force of the Larkin-type is generated.
This random 
force generates the Larkin-like roughness
\begin{align}
 \zeta_{\text{Larkin}} &= \varepsilon/2
\end{align}
which for the SDL with $\varepsilon=7$ 
implies a separate correlation length exponent
\begin{align}
 \nu_{\text{FS}} &= \frac{1}{4-7/2} = 2
\label{eq:nuFS=2}
\end{align}
according to the tilt-symmetry scaling relation (\ref{eq:nuscalrel}).

%%%%%%%%%%%%%%%%%%%%%%%%%%%%%%%%%%%%%
\subsection{Large force limit, crossover to single particle limit}
\label{sec:lf}

For sufficiently large external forces we can generalize the 
perturbative arguments for DLs from
Refs.\ \cite{Feigelmann1983,Sneddon1982} to
general elastic manifolds in $1+D$ Dimensions with an elastic energy
of the form (\ref{eq:ham}). 
Then, to second order in perturbation theory, 
the velocity of the center of mass of the line is
\begin{align}
 v
 &\approx F -\text{const}\times F^{\frac{{D}-2\elastsymbol}{2\elastsymbol}}
\end{align}
Here we assumed a short-ranged random potential 
that is completely uncorrelated along
the internal dimensions.
For the problem at hand ($D=1$, $\elastsymbol=2$) this implies that the first
correction at large forces should scale as
\begin{align}
 1-\frac{v}{F} &\sim F^{-7/4} \text{.}
  \label{eq:feigel}
\end{align}
This is in agreement with our numerical results (see Sec.\ \ref{sec:feigel})
 for
not too large forces. The asymptotic behavior for very large forces can be
understood with the same perturbative reasoning that led to eq.\
(\ref{eq:feigel}), but neglecting the elastic forces and considering the
effective single-particle (sp) equation $\frac{\partial u}{\partial t} =
\eta(u)+ F$. This leads to
\begin{align}
 v_{\text{sp}}&=F-\text{const}\times F^{-1}R_u^{-3} \\
 1-\frac{v_{\text{sp}}}{F} &\sim F^{-2}\quad  \text{.}
\end{align}
The crossover should happen, when the length scale $(\Delta_u/F)^{1/4}$ on which
the elastic adjustments to the forced induced motion are relevant becomes
significantly smaller than the lattice spacing $\Delta_x$. This is the
length scale that corresponds to the 
 time scale $\Delta_u/F$  for a moving
line to get to the next ``disorder site'' for  the free  dynamic 
exponent $z_0=4$.

%%%%%%%%%%%%%%%%%%%%%%%%%%%%%%
\subsection{Finite Temperature}

At finite temperatures $T>0$ there is a thermally activated motion, 
$\langle v\rangle>0$, for any driving force $F$. 
For the DL this dynamical phenomenon has successfully been described via the
thermal activation over barriers that are
determined from a static consideration as the motion is expected to
be very slow for low temperatures and forces
\cite{Chauve2000,Ioffe1987,Nattermann1990}.
For forces $F<F_T$ below depinning this involves activation 
over large energy barriers (diverging in the limit $F\approx 0$)
and results in so-called  creep motion. 
As a result of thermal activation, the sharp depinning transition 
at $F=F_T$ is rounded. 
For forces $F>F_T$ above depinning the line moves with finite 
velocity and additional thermal activation has only little effect.

For the SDL, there is an additional complication because 
of the   disorder-induced  localization 
transition at a
finite temperature $T_c$ \cite{Boltz2012,Boltz2013}.
For temperatures $T<T_c$, we expect the SDL to behave 
qualitatively similar to a  DL, i.e., to exhibit creep for $F<F_T$,
thermal rounding of the depinning transition at $F=F_T$, and 
only minor modifications of the flow behavior for $F>F_T$.
In order to derive the SDL creep law via a scaling argument, we
consider 
 the static equilibrium energy fluctuations, which  scale as
 $E_{eq} \sim L^{\omega}$
with the (equilibrium) energy fluctuation exponent
$\omega=D-2\elastsymbol+2\zeta_{eq}$. In a static framework, a depinning force
$F$ simply tilts the energy landscape
 $U_{F} \sim F L^D w \sim F L^{D+\zeta_{eq}} \text{.}$
Balancing these two contributions to optimize the total barrier energy
$E_{\text{barrier}}= E_{\text{eq}} - U_F$, one gets the energy of the
effective barriers scaling as
\begin{align}
 E_{\text{barrier}} &\sim F^{-\mu}
\end{align}
with the barrier exponent
\begin{align}
 \mu &= \frac{D-2\elastsymbol+2\zeta_{eq}}{2\elastsymbol-\zeta_{eq}} =
\frac{\omega}{2\elastsymbol - \zeta_{eq}}.
\end{align}
For lines with  $D=1$, this gives
 $\mu=1/4$ for the DL and $\mu\approx 0.07$ for the SDL. 
The velocity follows from the Arrhenius law to be
\begin{align}
 v&\sim \exp{\left[ -\text{const}\times F^{-\mu}/T \right]} \text{.}
\label{eq:arrhenius}
\end{align}
For the DL this 
has been confirmed experimentally \cite{Lemerle1998}.

For temperatures $T>T_c$, on the other hand, the scenario is less clear. 
In the
static problem, the SDL then depins already by thermal fluctuations. 
The roughness in the static
problem is only  larger than the thermal roughness, 
$\zeta_{eq}>\zeta_{th}$, for temperatures $T<T_c$
below the critical temperature \cite{Boltz2012,Boltz2013}. For 
$T>T_c$, the static SDL is thermally rough $\zeta_{th}=(4-D)/2$,
 and there are no
macroscopic energy fluctuations. Assuming that the static equilibrium physics
is  indeed relevant for low driving forces (as in the derivation of the creep
law), the conclusion could  be that there are
only finite energy barriers of 
 characteristic size $E_{\text{barrier}} = C$, and the 
velocity is given
by the so called thermally assisted flux flow (TAFF) 
\cite{Anderson1964,Kes1989}
\begin{align}
 v_{\text{TAFF}} &\propto F/T \exp{\left[-C/T\right]} \text{.}
\end{align}
The treatment within the FRG\ \cite{Chauve2000} suggests
that (at least to one-loop order) the force-force correlator $\Delta$ is only
affected by a finite temperature within the ``thermal boundary layer'' of width
$\sim T$ (especially, there is only a ``cusp'' 
for $T=0$). Within this layer the
line assumes the static roughness $\zeta_{eq}$, whereas on larger scales the
dynamic roughness $\zeta>\zeta_{eq}$ becomes apparent (for finite $v>0$). 
Thus, the FRG seems to be in line
with our previous reasoning, that is $\mu=0$ for $T>T_c$ and a TAFF-like
velocity-force curve. 

However, this comes with the substantial caveat that, to
our knowledge, the FRG theory in its present  form is not apt to
describe the full temperature dependence and, in particular,  
 the existence of a
transition to thermal roughness 
(which should manifest itself in the emergence
of a fixed point solution with roughness $\zeta_{th}$) at finite
temperature. One basic difficulty  is that a
 disorder-induced localization transition at a finite temperature $T_c$ 
does only occur for low dimensions $D<2\elastsymbol$, whereas 
 the FRG uses an expansion around an upper critical dimension 
$D=4\elastsymbol$. 
In section  \ref{sec:rounding}, we will present numerical evidence that 
the thermal rounding of the depinning transition at the threshold 
force  $F_T$ is 
very similar for DLs and SDLs. This surprising result 
suggests that the thermal depinning 
transition of  the SDL in the absence of a driving force
does {\em not} change  the depinning by a driving force qualitatively.

The numerical determination of the barrier exponent $\mu$ is an unsolved
problem even with algorithms specifically designed to capture the creep
dynamics \cite{Kolton2009}. The thermal rounding of the transition leads to a
temperature dependent velocity at the (zero-temperature) threshold force, 
which, for the DL, has been found to follow a power-law
\begin{align}
 v_{F_T} &= T^\psi. 
\label{eq:tr}
\end{align}
It has been
suggested that
$\psi=\beta/(1+2\beta)$ \cite{Chauve2000,TangUnpubl} (the
perturbative argument in Ref.\ \cite{Chauve2000} 
is equally valid for the SDL). 
Numerically and experimentally a value $\psi \approx 0.15
\approx \beta/(1+2\beta)$ ($\beta\approx 0.245$ \cite{Kolton2009})
has been  found for the DL with RB disorder 
\cite{Bustingorry2008,Bustingorry2012}.

%%%%%%%%%%%%%%%%%%%%%%%%%%%%%%%%%%%%%%%
\section{Numerical results} \label{sec:num}

%%%%%%%%%%%%%%%%%%%%%%%%%%%%%%%%%%%%%%%%
\subsection{Direct integration of the equation of motion}

We make use of a recently presented implementation \cite{Ferrero2013} for
graphics processing units \footnote{Our simulations were performed on a Tesla
C2070. We used the CUDA Toolkit, version
5.0.\cite{CUDA}} (GPUs).  The high number of parallelly executed
computations
becomes very advantageous for large lengths, with an effective speedup of two
orders of magnitude for the DL \cite{Ferrero2013}. 
As the
different elastic force generates only little additional branching (the
determination of the next-nearest neighbors with periodic boundaries), 
the GPU implementation 
is also  favorable for SDLs. 
 
Additionally, we implemented an equivalent
simulation for CPUs. An Euler integration scheme is used for the benefit of
computational simplicity.

In the numerical simulations we focus on random potential (RB) 
disorder (as opposed to Ref.\ \cite{Lee2006}). 
The random potential is implemented  by drawing random numbers
from a normal distribution on a $L\times M$ lattice where $M$ 
is the transverse size
of the system \cite{Rosso2002}. Between the lattice points 
the potential is interpolated by
periodic, cubic splines in $u$-direction.
Disorder averages were performed over 1000 samples.
Our system is periodic in $x$- and $u$-direction and
the simulation starts with a flat line $u(x)\equiv 0$.

We set the lattice spacing in both directions equal to one,
$\Delta_x=\Delta_u=1$, and approximate the fourth derivative by the central
finite difference
$\nabla^{(4)} u(x) \approx u(x-2)-4 u(x-1) + 6 u(x)-4 u(x+1)+u(x+2)$, which is
of second order in space. A von-Neumann stability analysis shows that (without
external forces) the Euler integration scheme becomes unstable for $\Delta
t/\Delta_x^2 > 2^{-3}$. Throughout this work we used 
time steps $\Delta t \leq 2^{-6}$,
unless stated differently.

As the width of
the line can be influenced by various effects on different length scales,
 it can
be difficult and error-prone to infer $\zeta$ directly from the 
line width $w$.
Therefore, it is helpful to study the structure factor
\begin{align}
 S(q,t) &= \overline{\lvert u(q,t)\rvert^2} \sim q^{-(2\zeta+1)}
\label{eq:sq}
\end{align}
where $u(q,t)$ is the Fourier transformation of $u(x,t)$. This assumes
self-affinity of the line.

\begin{figure}
 \includegraphics[width=0.95\linewidth]{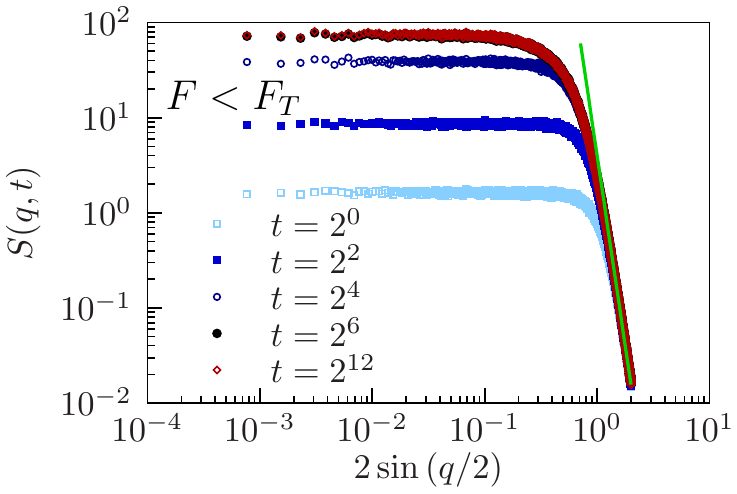}
 \caption{
Disorder averaged structure factor $S(q,t)$ for a SDL with $F=1.00 < F_T$
($L=8192$,
$M=16384$). The initial conditions (flat line)
persist on long length scales and cross
 over to a regime with Larkin roughness
($\zeta_{\text{Larkin}}=7/2$, solid line) on short length scales. 
After 
adjustment to the disorder till about $t=2^6$  the 
conformation of the line does not change.
}
 \label{fig:fltfc}
\end{figure}

For sufficiently large $M$ the system is ergodic only above the threshold, when
the line moves. Thus, for forces smaller than the threshold force 
(cp.\ Fig.\ \ref{fig:fltfc}) 
the line does not show the static roughness, but
adjusts itself to the potential on short length scales (large wavenumbers)
leading to Larkin roughness $\zeta_{\text{Larkin}}=7/2$, whereas the
conformation on longer length scales (small wavenumbers) 
depends on the initial
conditions. Here and in the following we plot the structure factor 
as a function
of $2\sin{(q/2)}$ to correct for lattice artifacts.

%%%%%%%%%%%%%%%%%%%%%%%%%%%%5
\subsection{Short-time dynamics scaling}

\begin{figure}
 \includegraphics[width=.95\linewidth]{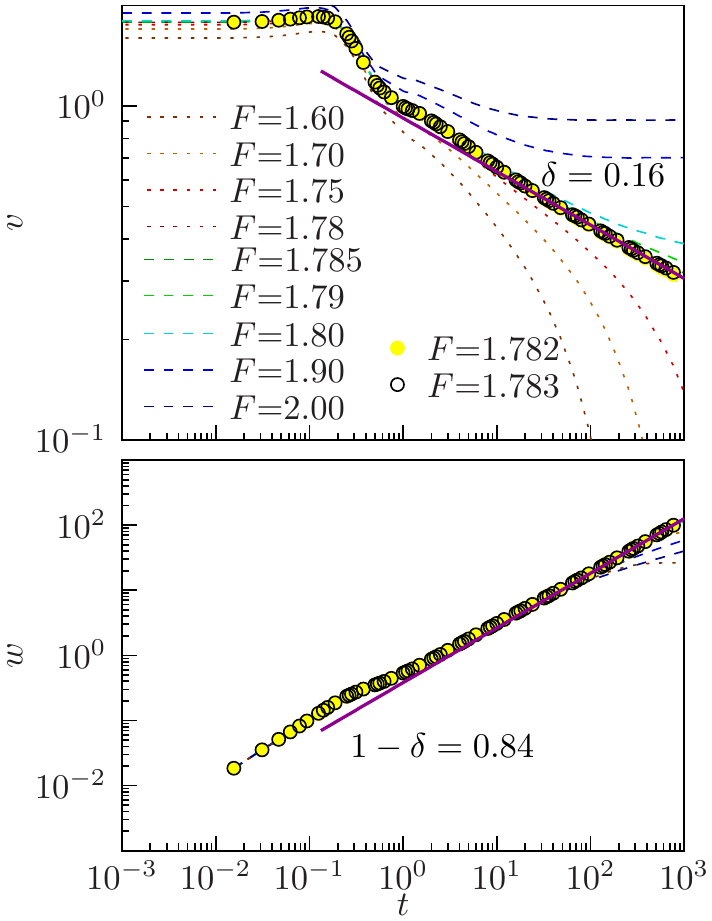}
\caption{
The disorder-averaged center-of-mass velocity $v(t)$ (top) and
roughness $w(t)$ (bottom) in a system with $L=8192$, $M=16384$ for various
forces and short times. 
Both quantities exhibit  power-law regime consistent 
with $\delta=0.16$ at $F\approx
1.782-1.783$. For clarity we plotted the data for these data as full
($F=1.782$) and hollow ($F=1.783$) circles and used dashed ($F>1.783$) and
dotted ($F<1.782$) lines for the rest (the key given in the upper figure holds
for both). We note that the
time-averaged roughness $\langle w(t)\rangle$ becomes maximal 
around $F\approx 1.782-1.783$. 
}
\label{fig:vcm1}
\end{figure}

Alike previous studies \cite{Kolton2006b,Ferrero2013, Lee2006} we employ
short-time dynamics scaling to determine the critical exponents of the SDL
at depinning. 
From eqs.\ (\ref{eq:scalingwtF}) and
(\ref{eq:vf})  we know that, at $F=F_T$, 
 velocity and line width scale as
\begin{align}
 v &\sim    t^{-\delta}   \sim t^{-\beta / \nu z} 
\label{eq:v}\\
 w &\sim t^{1-\delta} \sim t^{\zeta/ z} \quad \text{.}
\label{eq:w}
\end{align}
These two observables contain the same information regarding the critical
exponents.
The behavior for forces near the threshold can be used to extract
the exponent $\gamma=1/\nu z$ from rescaling the velocity according to
eq.\ (\ref{eq:vf}), 
\begin{align}
 v(t,F) t^{\beta / \nu z} \sim f_{\pm}(t^{1/(\nu z)} (F-F_T)). 
\label{eq:finscal}
\end{align}
However, with actual data it turns out to be difficult to extract precise
and unambiguous exponents from this finite-size scaling like procedure.
Additionally, the roughness exponent $\zeta$ becomes apparent in the structure
factor $S(q,t)$ for wavenumbers below some $q_t\sim L_t^{-1} \sim t^{-1/z}$,
see eq.\ (\ref{eq:sq}).

\begin{figure}
 \includegraphics[width=.95\linewidth]{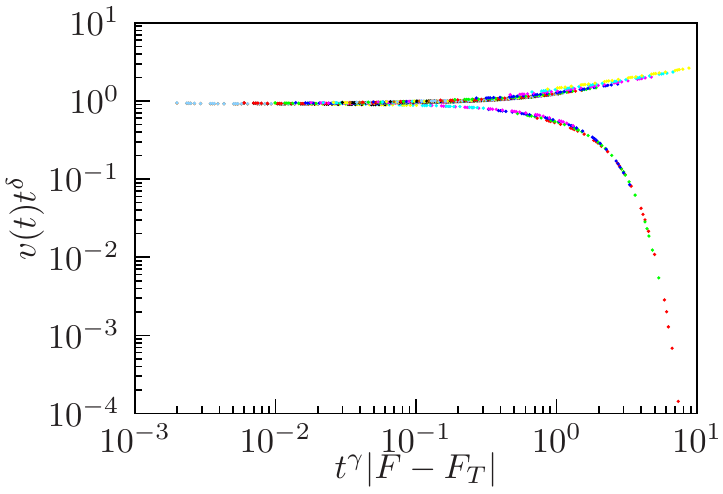}
\caption{
Scaling plot of the velocity for different forces and times using eq.\
\eqref{eq:finscal}. The upper branch contains forces $F>F_T$ and the lower one
forces $F<F_T$. Parameters are as for Fig.\ \ref{fig:vcm1}, but we used
more force values ($1<F<2$). As scaling is not expected to hold for
small times, we only considered $t>10$. In agreement
with Fig.\ \ref{fig:vcm1}, we used $F_T=1.7825$ and $\delta=0.16$. We see
satisfying scaling for $\gamma \approx 0.55$, 
which coincides with the findings
of Ref.\ \cite{Lee2006}.
}
\label{fig:finscal1}
\end{figure}

\begin{figure}
  \includegraphics[width=.95\linewidth]{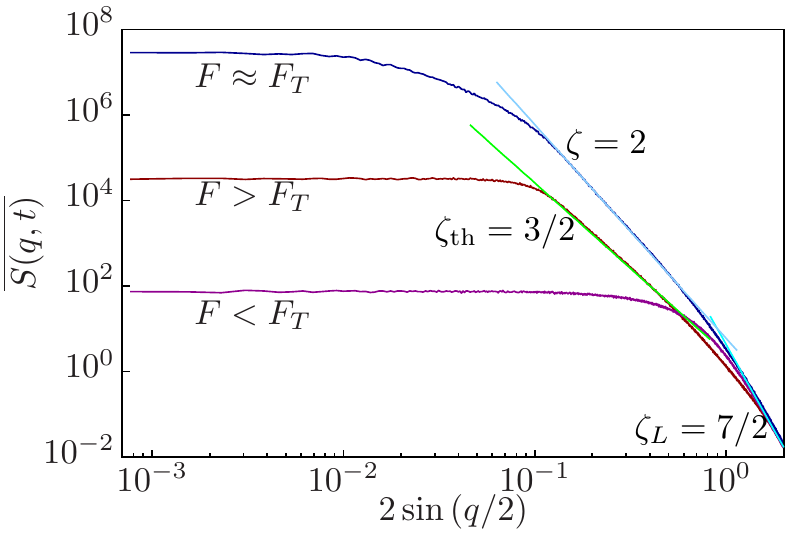}
\caption{
The disorder-averaged structure factor for a system with $L=8192$, $M=16384$ at
time $t=4096$ at three different forces representing the three regimes:
under-critical, critical and over-critical. 
Although configurations are not strictly self-affine 
(pure power-law behavior in the structure factor) there is a
clear emergence of an exponent $\zeta=2$ around the threshold force.
We identify this as the ``critical'' roughness exponent. As in 
Fig.\ \ref{fig:vcm1}, the system shows maximal
roughness at the threshold.
}
\label{fig:vergl}
\end{figure}

In Ref.\ \cite{Lee2006}, $\delta\approx 0.16$ has been found from studying
$v(t)$ and $w(t)$ for RF disorder. 
In Fig.\ \ref{fig:vcm1} we obtain the same  result for RB disorder. 
Additionally, we show in Fig.\ \ref{fig:finscal1} that the
data can be nicely matched using the scaling of eq.\ (\ref{eq:finscal}) and
$\gamma \approx 0.56$ in agreement with Ref.\ \cite{Lee2006}. By definition of
$\delta$ and $\gamma$, this implies $\beta=\delta/\gamma\approx 0.29$
and by means of scaling relations  $\zeta\approx 2.4-2.5$ as we already
pointed out in Sec.\ \ref{sec:scaling}. We do not find any evidence (at any
force) in the structure factor supporting this roughness. We do observe the
emergence of a new roughness exponent $\zeta \approx 2$ at higher forces
as can be seen from 
Fig.\ \ref{fig:vergl}. We consequently conclude that this is a more plausible
location of the threshold force and that $\zeta \approx 2$ is indeed the
threshold roughness exponent.

\begin{figure}
  \includegraphics[width=.95\linewidth]{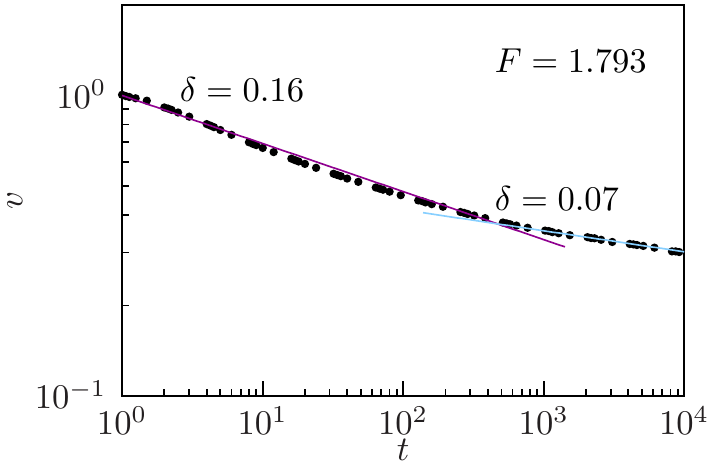}
\caption{
The disorder averaged center-of-mass velocity for $F=1.793$.
The system parameters are as for Fig.\ \ref{fig:vcm1}, 
which suggested $F_T\approx 1.782-1.783$. 
The description by a power-law fits 
better in the  ``macroscopic'' large time regime. 
Although
we rather determined $F_T$ by demanding a consistent $\delta$ 
as in Figs.\ \ref{fig:vcm1} and \ref{fig:finscal1}, we note 
that, for $F=1.793$, the second-power law fits particularly well.
}
\label{fig:v1793}
\end{figure}

The examination of $v(t)$ at higher forces and larger times reveals that
the curves that do not saturate to a finite $v$ for large times ($F>F_T$) 
or go to zero ($F<F_T$) seem to consist of two power-law segments, 
where only the first one for smaller times 
is consistent with $\delta\approx 0.16$, see Fig.\ \ref{fig:v1793}. 
This is analogous to the most recent 
findings for the DL in Ref.\ \cite{Ferrero2013}.
In Fig.\ \ref{fig:v1793} we show
$v(t)$ for $F=1.793$, which we believe to be close to the threshold 
force $F_T$ for the system size we use. 
As the exponent $\delta$ in the second power-law segment
is rather close to zero and we have no
independent method to determine $F_T$ (see also below in Sec.\
\ref{sec:middle}) giving a precise value for $\delta$ is difficult. As the
threshold roughness does not only influence the structure factor exactly at
$F_T$, but also for deviating forces (given that the correlation length is
still noticeable large), we think that we can rely on our value of
$\zeta\approx 2$ even though we do not know $F_T$ precisely. 
Additionally, we will  support the claim of a roughness exponent
$\zeta\approx 2$ with an independent method below
in Sec.\ \ref{sec:parabol}. 
In Fig.\ \ref{fig:v1793} we show that $F_T=1.793$ is consistent 
with an exponent value $\delta\approx 0.07$ for larger times $t$,
whereas $\delta\approx 0.16$ only at smaller times.

For $F_T=1.793$ we obtain, 
based on the numerical results $\delta\approx 0.07$ and $\zeta\approx 2.0$,
the following set of exponents:
\begin{align*}
 F_T &\approx 1.793~,~~&\delta &\approx 0.07~,\\
 \zeta &\approx 2.0~,~~& z &=\zeta/(1-\delta) \approx 2.15~,\\
 \nu &= \frac{1}{4-\zeta} \approx 0.5~,~~&\beta &=\nu(z-\zeta)\approx 0.08~,
\end{align*}
To determine $z$ we used the scaling relation (\ref{eq:delta}).
The value for $\nu$ then follows from the scaling relation 
 (\ref{eq:nuscalrel}) based on the tilt symmetry, the value 
for $\beta$ from the scaling relation (\ref{eq:delta}).

We note, however, that the value for $\beta$ is inconsistent 
with the value of $\beta\approx 0.29$  that we obtain numerically as 
explained in the next subsection.
A value  $\beta\approx 0.29$ implies 
  $\nu = \beta/(z-\zeta)\approx 1.92$ with errorbars that 
are consistent with the  exponent 
 $\nu_\text{FS}=2$ introduced above to characterize 
sample-to-sample fluctuations of the free energy, 
see eqs.\ (\ref{eq:nuFS}) and (\ref{eq:nuFS=2}).
Moreover, $\beta\approx 0.29$ is only consistent with 
the scaling relation  $\delta=\beta/\nu_\text{FS} z\approx 0.07$,
see eq.\   (\ref{eq:delta}), if we use 
$\nu_\text{FS}=2$. 
This suggests that the SDL in disorder is indeed characterized by 
two different exponents $\nu \neq \nu_\text{FS}$
 similar to charge density waves \cite{Middleton1992b}.
Therefore,  we conclude that all our data are best represented by the
following  extended set of exponents
(cp.\ Table \ref{tab:exp}):
\begin{align}
 F_T &\approx 1.793~,~~&\delta &\approx 0.07~, \nonumber\\ 
 \zeta &\approx 2.0~,~~ &z  &=\zeta/(1-\delta) \approx 2.15~,  \nonumber\\
 \nu &= \frac{1}{4-\zeta} \approx 0.5~,~~&\nu_\text{FS} &\approx 2~,
      \nonumber\\
\beta &\approx 0.29 \approx\nu_\text{FS}(z-\zeta).&&
   \label{eq:exponents}
\end{align}
We are not able to give sensible margins of errors and simulations of 
much larger systems (such that the ratio $w/vt$ becomes constant) 
might be needed.

%%%%%%%%%%%%%%%%%%%%%%%%%%%%%%%%%%%
\subsection{Velocity force relation}

\begin{figure}
 \includegraphics{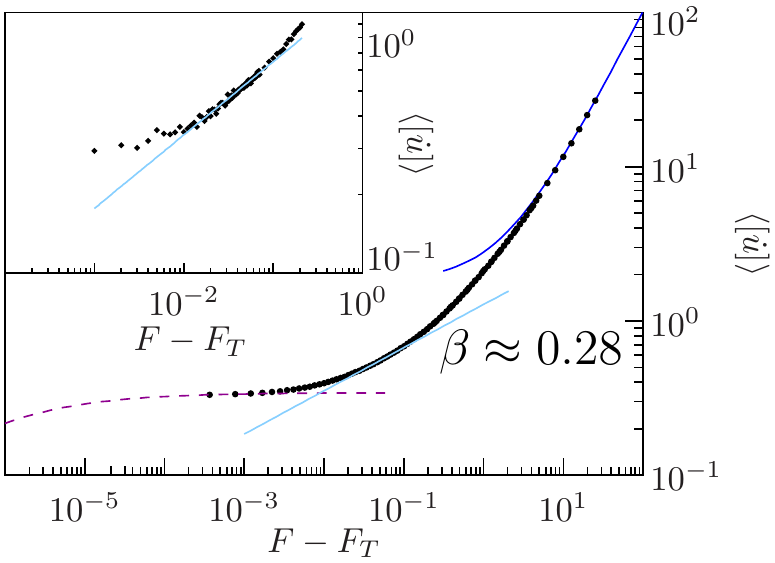}
 \caption{
Velocity force relation for a sample with of longitudinal size $L=2048$
and transverse size $M=8192$. The time average was computed over $10^{10}$
time steps $\Delta t = 2\cdot 10^{-2}$. 
We interpret the smaller slope at very low
force differences as the beginning of the ``single particle''-behavior
$v^{-1} \propto \text{const}+(F-F_T)^{-1/2}$ that has been derived in Ref.\
\cite{Duemmer2005}, whereas at higher force difference the moving line with
$v\sim F$ becomes apparent. The best value of $\beta$ depends on
the threshold force used (here and in the inset we used $F_T\approx1.79$), but
we found consistent values of $\beta\approx 0.28$ for various sizes and
samples. Inset: velocity force relation for one larger sample 
with $L=8192$ and $M=16384$.
The time average was computed over $2^{20}$ time steps 
$\Delta t = 2^{-6}$. There is
a larger window with visible scaling.
} 
\label{fig:beta}
\end{figure}

We also tried to determine the velocity exponent $\beta$ directly. In the
treatment of DLs, there is a significant discrepancy in the reported values of
$\beta$, with either $\beta\approx 0.33$ \cite{Duemmer2005} or $\beta\approx
0.25$ \cite{Ferrero2013}. These were determined by means of two slightly
different approaches: one can either determine 
(as in \cite{Duemmer2005}) the
threshold force $F_{T,\text{sample}}$ for each sample and average
$\overline{\langle [\dot{u}] \rangle(F-F_{T,\text{sample}})}$ or use
$\langle v\rangle(F-\overline{F_{T,\text{sample}}})$ (as in
\cite{Ferrero2013}). We chose the first approach because
$\overline{F_{T,\text{sample}}}$ cannot be clearly extracted from short-time
dynamics scaling and, thus, the sample specific threshold force has to be
determined anyway.

In Fig.\ \ref{fig:beta} we show
 data for two single samples of sizes $L=2048$ and $L=8192$. We
infer from our data
\begin{equation}
 \beta \approx 0.28 \text{.}
\end{equation}
This agrees with the value $\beta \approx 0.29$ found in Ref.\
\cite{Lee2006}.

%%%%%%%%%%%%%%%%%%%%%%%%%%%%%%%%
\subsection{Large forces}
\label{sec:feigel}

\begin{figure}
 \includegraphics[width=.95\linewidth]{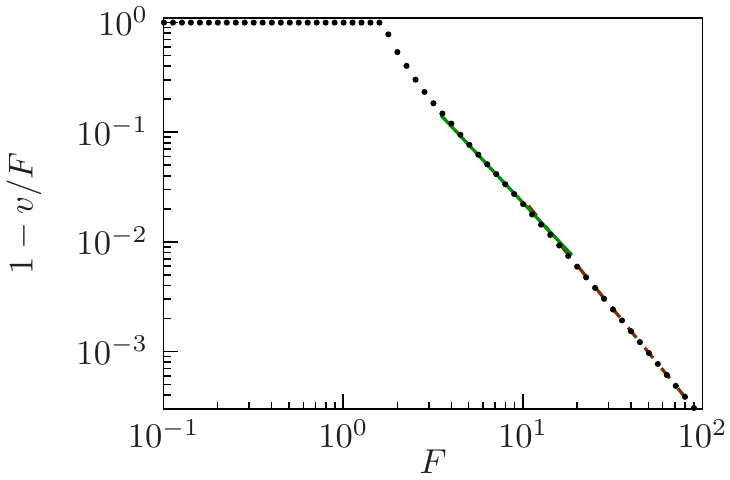}
 \caption{
The deviation from the asymptotic $v\sim F$ for large forces in 
\emph{one} sample. The green solid line is the perturbative result
as given by  eq.\
(\ref{eq:feigel}), the brown dashed line is given by $1-v/F\sim F^{-2}$. A
similar cross-over to the single-particle behavior has been observed for the
DL \cite{Dong1993}. We used an Euler integration scheme with $2^{20}$ small 
time steps $\Delta t=2^{-10}$ to accurately simulate
the system even for large forces. For steady state results the
time average was performed only over the last $2^{19}$ time steps. The system
 has lateral size $L=64$ and transverse size $M=512$.
}
\label{fig:feigel}
\end{figure}

We have successfully confirmed the perturbative results 
of Sec.\ \ref{sec:lf},
cp.\ Fig.\ \ref{fig:feigel}, for large driving forces. 
Additionally, we checked that the roughness
exponent $\zeta$ of the SDL 
takes its thermal value $\zeta_{th} = 3/2$
for sufficiently large driving forces. 

\begin{figure}
 \includegraphics[width=.95\linewidth]{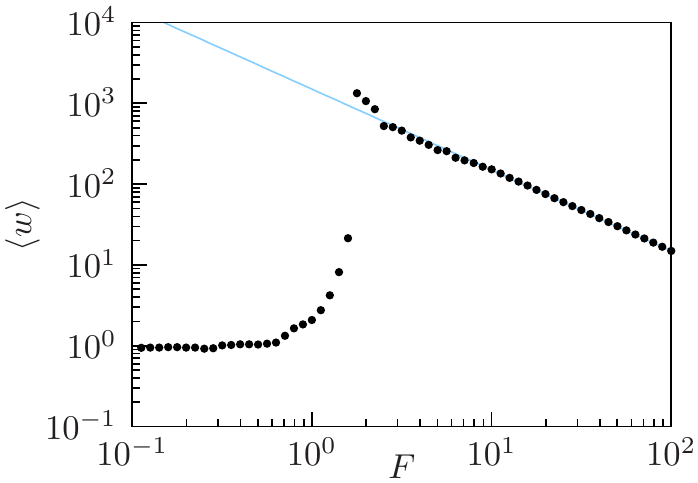}
 \caption{
Time averaged width for a sample with $L=M=512$. The time average was
performed over $2^{30}$ time steps. The line is maximally rough at the
threshold, which is consistent with a ``shockwave'' motion, where small
segments of the line move whilst the rest is blocked. At higher forces the
width reduces with the force, which could be interpreted as a decrease of the
effective temperature or, analogously, an increase in the effective persistence
length. 
The solid line is a 
guide to the eye showing the $F^{-1}$ trend of the width.
}
\label{fig:rauhvsf}
\end{figure}

An interesting (and maybe
counterintuitive) result is that increasing the force leads to a decrease in
the ``effective'' temperature. 
We show this in Fig.\ \ref{fig:rauhvsf} using
the time averaged width (cp. eq.\ \eqref{eq:defw}) of the line 
as function of the force. 
We expect the width $w$  to scale as $w^2 \sim L_p^2 (L/L_p)^{2\zeta}$ 
with an effective persistence length $L_p$ of the SDL.
For purely thermal fluctuations we have $\zeta=\zeta_{th}=3/2$ and 
$L_p \sim  1/T$ \cite{Kleinert2006,Gutjahr2006}.
 For 
 the static SDL in disorder, on the other hand, 
 we found a disorder-induced persistence length
 which is independent of temperature  in the
 low-temperature phase of the SDL and which is minimal at the delocalization 
threshold
\cite{Boltz2012,Boltz2013}.
For the dynamic depinning, on the other hand, 
the simulation results in  Fig.\ \ref{fig:rauhvsf} show an 
increase of the SDL width at depinning, followed by a
decrease $w\sim F^{-1}$ at high forces, where the roughness exponent 
assumes its thermal value $\zeta=\zeta_{th}=3/2$ again. 
This is consistent with a behavior $w^2 \sim L^3/L_p$ with 
$L_p \sim F$ and, thus, an ``effective'' temperature $T_{\rm eff} \sim
F^{-1}$ that {\em decreases} with driving force $F$. 
This can be rationalized
from the FRG approach by noting that for high velocities the
disorder contributes in form of an additional thermal noise corresponding to a
temperature $T_\text{eff} \sim v \int \Delta$ \cite{Chauve2000}. 
At high forces the line is far
from the threshold and, thus, $\int \Delta$ should be close to the RB-disorder
value of $\int \Delta = 0$.
The observed behavior $T_{\rm eff} \sim F^{-1}$ 
implies $\int \Delta \sim F^{-2}$ because
$v\sim F$.

%%%%%%%%%%%%%%%%%%
\subsection{Confinement in a moving parabolic potential}
\label{sec:parabol}

A different approach \cite{LeDoussal2006,LeDoussal2007,Middleton2007,Rosso2007}
to compute the threshold force and, additionally, the effective disorder
correlation functions is to pull the line very slowly with a spring.
This means to introduce a parabolic potential 
acting on each line segment according to
\begin{align}
 V_{\text{par}}(u,t) &= m^2 (u-w(t))^2
\end{align}
and move 
the center $w(t)$ of the parabolic potential 
moves with a (small) constant velocity
$\dot{w}=\text{const}>0$. We call $m^2$ the strength of the potential.
The underlying idea is essentially 
that the force $F(\dot{w})$ exerted by  the parabolic potential 
on the line as it moves forward  becomes the threshold force for
$m\rightarrow 0$, $\dot{w}\rightarrow 0$. 
More precisely,  it was found for the DL that
\begin{align*}
 \langle w - [u(w)]\rangle m^2 = F_T + C m^{2-\zeta}
\end{align*}
with some constant $C$ (from now on, we assume that $\dot{w}$ is sufficiently
small). 
For general $\elastsymbol$, 
the expected corrections due to finite values of $m$ have to
be adjusted to account for the different elasticity.  The length scale $L_m$ at
which the confinement through the parabolic potential becomes relevant follows
from balancing the elastic and the potential energy per length
\begin{align*}
 u^2/L^{2\elastsymbol} \sim m^2 u^2\\
 L_m^{\elastsymbol} \sim m^{-1}
\end{align*}
$L_m$ is the length of independently adjusting line-segments. 
Averaging over all
$w$, i.e., averaging over disorder, typical displacements
scale as $u_m \sim L_m^\zeta$. This leads to effective forces scaling as 
\begin{align}
 F_m &\sim m^2 u_m \sim m^2 L_m^\zeta\\
     &\sim m^{2-\zeta/\elastsymbol},
\end{align}
which includes the aforementioned DL result ($\elastsymbol=1$). A different
approach \cite{Ferrero2013} leading to the same result would 
be to use the known
scaling of the finite-size corrections to the threshold force in one sample
$F_T(L)-F_T\sim L^{-1/\nu}$ 
together with the notion that the relevant
length scale is imposed by the parabolic potential and therefore given by
$L_m\sim m^{-1/\elastsymbol}$. The confinement splits the line into independent
segments of length $L_m$ and, therefore, one gets 
$F_m = F_T (L_M)-F_m m^{1/\nu \elastsymbol}$. 
Using eq.\ \eqref{eq:nuscalrel} one finds 
$1/(\nu \elastsymbol) = 2- \zeta/\elastsymbol$ and, thus,
 these two approaches are equivalent. In this derivation
it is also clear that there will be deviations for very small $\elastsymbol$
when
$L_m$ exceeds $L$. 

\begin{figure}
 \includegraphics[width=.95\linewidth]{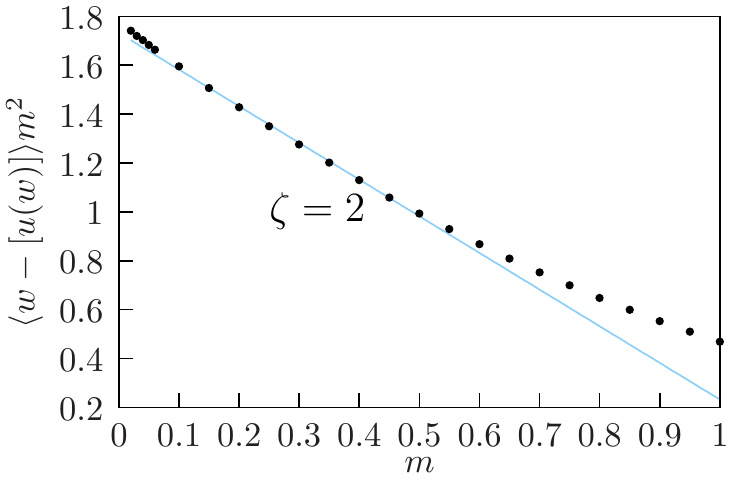}
 \caption{
Time average of the net force that the parabolic potential exerts on the
center of motion of the line ($L=256$, $M=1024$). The solid line is a fit for
$m\in [0.1,0.5]$ with $\zeta=2$ yielding $F_T=1.733\pm 0.007$. The
velocity of the parabolic potential was $\dot{w}=10^{-6}$. The shape of the
numerical results and their deviations from the analytical expectation resemble
the findings for the DL in Ref.\ \cite{Middleton2007}. The value $F_T=1.733$
deviates from the value found above
because of the differing system size.
}
\label{fig:parabol}
\end{figure}

As the line is constantly moved forward we chose to change the implementation
of the potential. We still have a fixed amount $M$ of potential values (knots
for the cubic spline), but we update the potential ``on-the-fly'' as the line
is moved forward. Every time a segment of the line  reaches a new quarter of
$M$, we update the quarter that has the greatest distance to the current
location and compute the splines  
[in principle, this changes the spline at the
current location of the line, but the change is negligibly small if $M$ is large
enough (we used $M=1024$)]. 
Our motivation for this scheme was to avoid
finite-size-effects in the transverse direction and to be able to compute the
disorder average as a time average. In Fig.\
\ref{fig:parabol} we show that our data are consistent with 
$\zeta\approx 2.0$.

\begin{figure}
 \includegraphics[width=.95\linewidth]{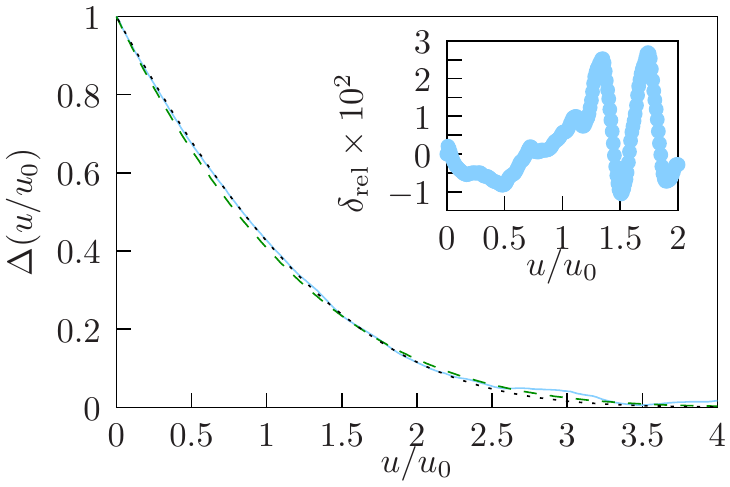}
 \caption{
Numerical determination of the disorder correlator $\Delta(u)$ 
according to (\ref{eq:Deltaunum}) with 
$m^2=10^{-2}$. The data are taken from the same simulation as for
Fig.\ \ref{fig:parabol}. The dashed (green) line is the one-loop
solution and the dotted (black) line is the two-loop solution, see Sec.\
\ref{sec:frg} and Ref.\ \cite{LeDoussal2002}. The length scale $u_0$ is
determined via $\int \mathrm{d}u \Delta{(u/u_0)} = 1$. The inset shows the
relative
difference between the two-loop solution $\Delta_{2L}$ and the numerically
determined values $\Delta$, i.e.,
$\delta_{\text{rel}}=(\Delta-\Delta_{2L})/\Delta_{2L}$ for small $u/u_0$.
}
\label{fig:parabolkumul}
\end{figure}

Furthermore, this setup allows for a direct measurement of the effective force
correlation function and hence a validation of the renormalization group
solution. This is achieved via the second cumulant of $u(t)-w(t)$ as
\begin{equation}
 \langle (u(t) - w(t)) (u(t')-w'(t'))\rangle_c m^4 \propto 
    \Delta(w-w') \text{.}
\label{eq:Deltaunum}
\end{equation}
This contribution is closely related to the shape
of the force correlation function \cite{LeDoussal2002}, 
see eq.\ (\ref{eq:zeta}).
The existence of a non-vanishing $\Delta'(0^+)$, often referred to as
``cusp'', is a sign of the non-analyticity of the correlation function. 
We show our results in Fig.\ \ref{fig:parabolkumul}, 
which demonstrate
promising agreement with the functional renormalization group fixed point
function.

%%%%%%%%%%%%%%%%%%%%%%%%%%%%
\subsection{Finite temperatures -- thermal rounding}
\label{sec:rounding}

The finite mean velocity at arbitrarily small non-zero
temperatures is only visible for very long simulation times, 
Thus, the regime in which creep/TAFF
behavior should occur is not accessible for us.

\begin{figure}
 \includegraphics[width=.95\linewidth]{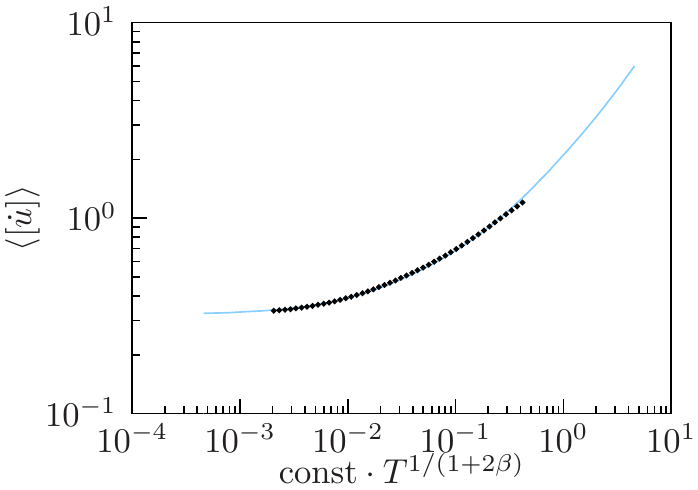}
 \caption{
Time average of the velocity at the threshold force for finite
temperatures. We used $\beta=0.28$ or $\eta=0.64$ in the rescaling of the
temperature (black points). 
The data collapse with data for the velocity as a function of 
$F-F_T$ at $T=0$ (blue line) 
after rescaling using eq.\ (\ref{eq:rescale}). 
The data for  the velocity as a function of 
$F-F_T$ at $T=0$  and, thus, all system parameters but
the temperature are the same as for Fig.\ \ref{fig:beta}.
}
\label{fig:thermalround}
\end{figure}

The thermal rounding exponent $\psi$, as defined in eq.\ \eqref{eq:tr}, can be
interpreted in a slightly different way, that we feel is more apt for the
interpretation of numerical data. Moving away from the threshold at 
$F=F_T$ and $T=0$
the velocity scales as $(F-F_T)^\beta$ or $T^\psi$, respectively. Thus, a
finite temperature at $F_T$ can effectively be seen as a contribution to the
pulling force with
\begin{equation}
 F-F_T \sim T^\eta
\label{eq:rescale}
\end{equation}
and $\eta=\psi/\beta$. Adapting our previous statements the perturbative
conjecture would be $\eta=1/(1+2\beta)$. In Fig.\ \ref{fig:thermalround} we
show that using this rescaling we can collapse data for the 
velocity as a function of $F-F_T$ at $T=0$ and 
for the velocity at the threshold 
$F=F_T$ as a function fo temperature. 
The data collapse is  consistent
with $\eta=1/(1+2\beta)$ and $\beta\approx 0.28$.

\begin{figure}
 \includegraphics[width=.95\linewidth]{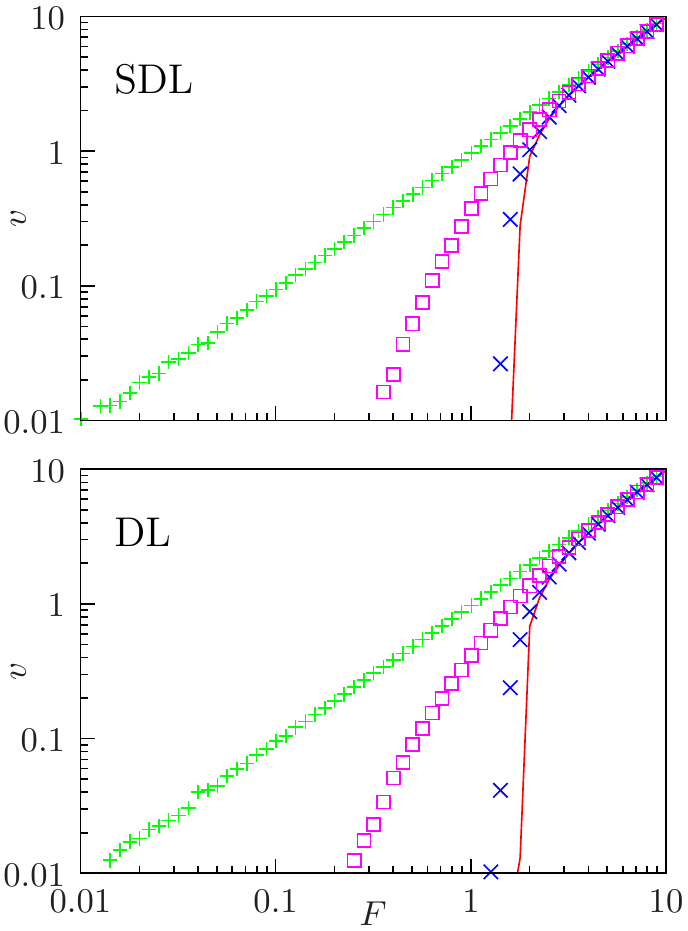}
 \caption{
Velocity of the line as a function of the driving force at four
different temperatures (solid line: $T=0$). For a better comparison we show data
for a SDL and a DL in the same disorder. Note that as the same disorder was used
the velocities are very similar, but not identical. All system parameters but
the temperature are the same as for Fig.\ \ref{fig:beta}.}
\label{fig:vf_comb}
\end{figure}

We compare our numerical results for 
the thermal rounding of the depinning transition 
for the DL and the SDL in Fig.\ \ref{fig:vf_comb}.
Surprisingly, 
we find {\em no} evidence  for a qualitative change
at a finite temperature that could be associated with the localization
transition in the static problem. This could either mean that the change is too
subtle to be apparent within our numerics or that, in terms of the FRG,
 a finite velocity $v>0$ implies that the relevant fixed point 
is one featuring $v>0$ and 
$T=0$, which would make the transition at finite temperature irrelevant for a
moving line.

%%%%%%%%%%%%%%%%%%%%%%%%%%%%%%%%%%%%%%%%%%%%%%%%%%
\section{Direct computation of threshold force -- 
  Middleton's theorems}
\label{sec:middle}

In the study of the depinning of directed lines two important 
properties have
been found \cite{Middleton1992}: 
a) the ``no-passing'' theorem, which
states that two lines in the same disorder realization that do not cross each
other at a given time will never cross each other; 
b) if each segment of the
line has at some point in time a non-negative velocity the velocity will
remain
non-negative for all times; this has been referred to as ``no-return''
property \cite{Kolton2006}.
The no-passing theorem is believed to hold for every
convex next-neighbor elastic energy. The combination of these
properties 
allows for a fast and precise algorithm to determine
the threshold force directly \cite{Rosso2002}. 

\begin{figure}
 \includegraphics[width=.8\linewidth]{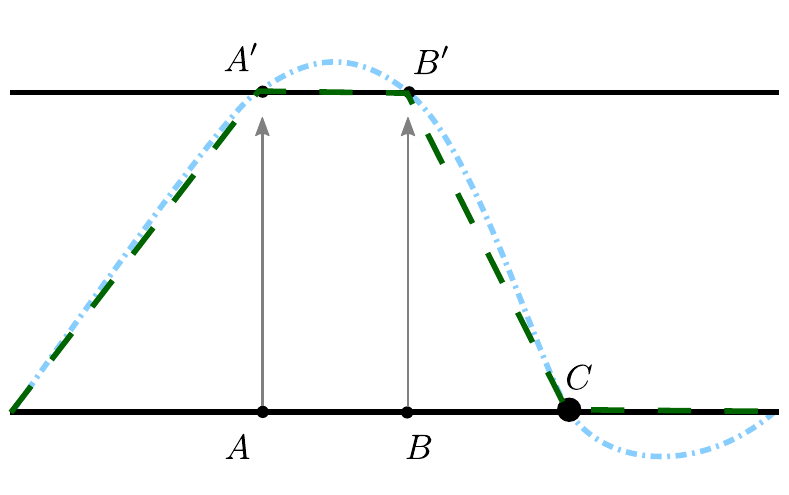}
 \caption{
Cartoon exemplifying the difference in the evolution of a DL and
a SDL starting from the lower horizontal conformation. Let the potential be
such, that the line points at $A$ and $B$ are forced to move to $A'$ and $B'$,
whereas the ends and the line point at $C$ cannot move. The DL (green dashed
line) does not pass the blocked upper line and also does not cross the lower
line. The SDL (blue dash-dotted line) seeks to reduce the curvature and,
consequently, does cross both lines.
}
\label{fig:diffdlsdl}
\end{figure}

With regard to the no-passing theorem 
we consider the following situation: 
two lines $z_1$, $z_2$ in the same medium that touch each 
other at exactly one
point, that is there is one $x$ with $z_1(x)=z_2(x)$ and $z_1(x') >
z_2(x')$ for $x\neq x'$. For the DL, it follows that $v_1(x)>v_2(x)$. As both
lines touch each other in $x$ a possibly differing velocity of the two lines is
due to the elastic forces. In the discrete version we then have 
$v_1(x)>v_2(x)$, because $(z_1(x-1)-z_2(x-1))+(z_1(x+1)-z_2(x+1))>0$.  This
does not work for the SDL, because the difference in velocity
\begin{align}
v_1(x)-v_2(x)=-&(z_1(x-2)-z_2(x-2))\nonumber\\
-&(z_1(x+2)-z_2(x+2))\nonumber \\
+4&(z_1(x-1)-z_2(x-1))\nonumber\\
+ 4 &( z _1(x+1)-z_2(x+1))
\end{align}
can take any value. We visualize this in Fig.\ \ref{fig:diffdlsdl}. 

Therefore, 
the SDL  does not necessarily explore ``new'' regions of the disorder
potential. A line that moves back and forth ``knows'' essentially the whole
potential at any point. This could 
be an explanation for the two distinct correlation 
length exponents $\nu$ and  $\nu_\text{FS}$ that we found for 
the SDL, see eq.\ (\ref{eq:exponents}). 
The fact that Middleton's theorems
guarantee that $u(x)$ is a monotonic function of time is also used in the
evaluation of ambiguous vertices within the FRG treatment in Ref.\
\cite{LeDoussal2002}. However, we feel that coming from a moving line with
$v>0$ the quasistatic depinning limit is still well defined for the SDL,
because the long time limit of the roughness is (for finite $L$) finite and,
therefore, all segments will eventually move on average with the same velocity.
The agreement of the FRG with our numerical results supports that there is no
fundamental problem with the applicability of the FRG for the SDL. 
Still, there
is definitely room and need for a more rigorous analysis.

%%%%%%%%%%%%%%%%%%%%%%%%%%%%%%%%%%
\section{Conclusion} 
\label{sec:concl}

We studied the depinning of SDLs  from  disorder
(RF or RB) in $1+1$ dimensions due to a 
driving force.
Using scaling arguments, analytical FRG calculations, and extensive 
numerical simulations, we characterized the critical behavior
at and around the depinning transition. 
Our study revealed some characteristic differences in depinning behavior 
between SDLs and DLs governed by tension. 

 The  resulting equation of motion for the SDL in disorder 
 is equivalent 
to the Herring-Mullins equation for surface growth, which is 
 governed by surface
diffusion rather than surface tension, in quenched disorder. 
Our results also apply to semiflexible polymers with contour lengths 
smaller than their persistence length,
 which are pulled over
 a disordered surfaces or driven through a random medium.

We show that Middleton's theorems do not apply to SDLs.
Nevertheless we find a well-defined threshold force $F_T$ for 
depinning. 
Likewise, critical exponents characterizing roughness and the dynamics 
of depinning can be defined and numerically 
determined for the SDL as for the DL. The SDL represents an own
dynamical universality class with a different set of exponents. 
Our extensive numerical data is best described by the set 
(\ref{eq:exponents}) of critical exponents which is also consistent 
with scaling relations, see section \ref{sec:scaling}.
We also investigated the behavior of the SDL persistence length, which 
exhibits a characteristic non-monotonous 
force-dependence through the depinning transition
(see Fig.\ \ref{fig:rauhvsf}).

 We transferred functional renormalization group results to the
elasticity of stiff interfaces, which allows us to 
derive analytical results or bounds 
for critical exponents (see section \ref{sec:frg}).
We find satisfying agreement of these analytical predictions with
our numerical work. 
Our results indirectly imply that the depinning threshold is
associated with two distinct correlation length exponents
$\nu$ and $\nu_\text{FS}$. To our knowledge
this would be the first occurrence of such behavior in a non-periodic system. 
This result could be linked to the non-validity of 
Middleton's no-passing theorem.

Our findings for the critical exponents at the 
threshold force disagree in parts
with previous numerical work, which suggests that further work, 
especially on
much larger systems, might be helpful to settle these exponents.

For finite temperatures, the depinning of a SDL 
is an interesting problem because, at equilibrium (no pulling force),
the  problem features a
disorder-driven localization transition at a finite temperature. 
Such a transition is absent for the DL, which remains in a localized
disorder-dominated phase  for all temperatures. 
Surprisingly, the numerical results for a comparison of the 
thermal rounding of the force-driven depinning transition do not 
show any qualitative difference between DLs and SDLs, see 
 Fig.\ \ref{fig:vf_comb}.
In a renormalization group sense, 
this might imply that the force-driven depinning and the temperature-driven 
delocalization are not described by the same fixed point.

%%%%%%%%%%%%%%%%%%%%
\section{Acknowledgments.}
We acknowledge financial support by  the Deutsche Forschungsgemeinschaft
(KI 662/2-1).

\appendix
\section{Bounds for the dynamical exponent $z$}
\label{app:bounds}

In Ref.\ \cite{LeDoussal2002}  the dynamical exponent $z$ has
been found to be to two-loop order
\begin{align}
 z &= 4 - 2/9 \varepsilon + \varepsilon^2 \left( \frac{\zeta_2}{3} - \frac{2
X^{(2\elastsymbol)}}{27} + \frac{Y^{(2\elastsymbol)}}{54} \right)
\end{align}
with $\zeta_2$, $X^{(2\elastsymbol)}$ the same as in the main text and
\begin{align}
 Y^{(2\elastsymbol)} &= X^{(2\elastsymbol)} + \varepsilon \left( \frac{2
I_\eta}{(\varepsilon I_1)^2} -
\frac{1}{\varepsilon} \right) \text{,}
\end{align}
where we use $I_\eta$ as shorthand notation for the aforementioned correction
to friction for the SDL
\begin{align}
I_\eta=I_\eta^{(4)}&=\int_{q_1,q_2}
\frac{1}{
\left(q_1^2+m^2\right)^2\left(q_2^2+m^2\right)^4} \times\\
&\frac{1}{
\left(\left(q_2^2+m^2\right)^2+
\left(\left(q _1-q_2 \right)^2+m ^2\right)^2 \right)}
\end{align}
 and
$I_1$ is the one-loop
integral given by
\begin{align}
 I_1 &= m^{-\varepsilon} \frac{\Gamma(\varepsilon/2)}{\Gamma(4)} \left(\int_q
\mathrm{e}^{-q^2}\right) \text{.}
\end{align}
Using $I_\eta \geq 0$ we find the lower bound
\begin{align}
z \geq 1.8607 \text{.}
\end{align}
For an upper bound we note that in any dimension (any value of $\epsilon$)
the following inequality holds
\begin{align*}
  I_\eta \leq J_\eta &= \int_{q_1,q_2}
\frac{1}{(q_1^2+m^2)^2(q_2^2+m^2)^4(q_3^2+m^2)^2} 
\end{align*}
and $J_\eta$ can be evaluated to leading order in $\epsilon$ via Laplace
transforms
\begin{align*}
 J_\eta &= \frac{1}{\Gamma(4)} \int\limits_{\substack{s_1,s_2,s_3>0\\q_1,q_2}} s_1
s_2^3 s_3 \times \\
&~~\times
\mathrm{e}^{-s_1(q_1^2+m^2)-s_2(q_2^2+m^2)-s_3(q_3^2+m^2)} \\
&= \frac{\left(\int_q
\mathrm{e}^{-q^2}\right)^2}{\Gamma(4)}\int\limits_{s_1,s_2,s_3>0} \frac{s_1
s_2^3 s_3 \mathrm{e}^{-m^2(s_1+s_2+s_3)}}{(s_1
s_2 +s_2 s_3+s_2 s_3)^{d/2}} 
\end{align*}
Substituting $s_2
\rightarrow s_1 s_2$, $s_3\rightarrow s_1 s_3$ gives
\begin{align*}
 J_\eta &= \frac{\left(\int_q \mathrm{e}^{-q^2}\right)^2}{\Gamma(4)}
\int\limits_{s_1,s_2,s_3>0}\frac{s_1^{7-d} s_2^3 s_3 \mathrm{e}^{-m^2 s_1
(1+s_2+s_3)}}{(s_2+s_3+s_2 s_3)^{d/2}}
\nonumber\\
 &= \frac{\left(\int_q \mathrm{e}^{-q^2}\right)^2}{\Gamma(4)}
\Gamma(\varepsilon)
m^{-2\varepsilon} \times\nonumber\\
&~~\times\int\limits_{s_2,s_3>0}\frac{s_2^3s_3}{(s_2+s_3+s_2 s_3)^{d/2}}
\frac{1}{(1+s_2+s_3)^\varepsilon}\nonumber\\
&=\frac{\left(\int_q \mathrm{e}^{-q^2}\right)^2}{\Gamma(4)} 
\Gamma(\varepsilon)m^{-2\varepsilon} J \nonumber
\end{align*}
\begin{align*}
 J &= \int\limits_{s_2,s_3>0} \frac{s_2^3 s_3}{(s_2+s_3+s_2 s_3)^4}
\frac{(s_2+s_3+s_2
s_3)^{\varepsilon/2}}{(1+s_2+s_3)^\varepsilon}\nonumber\\
  &= \int\limits_{s_2,s_3>0} \frac{s_2 s_3}{(1+s_3+s_2 s_3)^4}
\frac{s_2^{\varepsilon/2} (1+s_3+
s_2 s_3)^{\varepsilon/2}}{(1+s_2+s_2 s_3)^\varepsilon}\nonumber\\
&= J_1+J_2+J_3
\end{align*}
In the second to last step we substituted $s_3\rightarrow s_2 s_3$. We have
divided the integration in three terms to isolate the (important) divergent
part.
\begin{align}
 J_1 &= \int_{1}^\infty \mathrm{d}s_2 \int_{0}^\infty \mathrm{d}s_3  \frac{s_2
s_3}{(1+s_3+s_2 s_3)^4} \times \nonumber\\
&~~\times\frac{s_2^{\varepsilon/2} (1+s_3+
s_2 s_3)^{\varepsilon/2}}{(1+s_2+s_2 s_3)^\varepsilon} \nonumber\\
 &= -\frac{1}{12}+ \frac{\ln{2}}{6} + \mathcal{O}(\varepsilon)
\end{align}
\begin{align}
J_2 &= \int_0^\infty \mathrm{d}s_3 \int_0^1 \mathrm{d}s_2 \left[ \frac{s_2
s_3}{(1+s_3+s_2 s_3)^4} \times \right. \nonumber\\
&\left. ~~\times
\frac{s_2^{\varepsilon/2} (1+s_3+
s_2 s_3)^{\varepsilon/2}}{(1+s_2+s_2 s_3)^\varepsilon} -
\frac{s_2^{1+\varepsilon/2} s_3}{(1+s_2 s_3)^{4+\varepsilon/2}} \right] 
  \nonumber\\
&= -\frac{1}{12}-\frac{\ln{4}}{6} + \mathcal{O}(\varepsilon)
\end{align}
\begin{align}
J_3 &= \int_0^\infty \mathrm{d}s_3 \int_0^1 \mathrm{d}s_2
\frac{s_2^{1+\varepsilon/2} s_3}{(1+s_2 s_3)^{4+\varepsilon/2}} 
 \nonumber\\
&= \frac{1}{3 \varepsilon} - \frac{5}{18}
\end{align}
Thus collecting all terms we find
\begin{align}
 J_\eta &= \frac{\left(\int_q \mathrm{e}^{-q^2}\right)^2}{\Gamma(4)} 
\Gamma(\varepsilon)m^{-2\varepsilon} (J_1+J_2+J_3) 
\nonumber\\
&= \frac{6}{4} (J_1+J_2+J_3) (\varepsilon I_1)^2
\end{align}
and 
\begin{align}
Y^{(4)}&\leq X^{(4)} + \varepsilon \left( \frac{2 J_\eta}{(\varepsilon I_1)^2} -
\frac{1}{\varepsilon} \right)\nonumber\\
  &\leq -\frac{2}{3}+ \mathcal{O}(\varepsilon) \text{,}
\end{align} 
which ultimately yields
\begin{align}
 z &\leq 2.3144,
\end{align}
i.e., eq.\ (\ref{eq:zbounds}) in the main text.

%%%%%%%%%%%%%%%%%%%%%%%%%%%%%%%%%%%%%%%%%%%%%%%%%%%%%%%%%%%%%%%%%%

%

\end{document}